\documentclass[preprint,3p,times]{elsarticle}

\usepackage[utf8]{inputenc}
\usepackage[T1]{fontenc} 
\usepackage{array} 
\usepackage{hyperref} 
\usepackage{lineno} 
\modulolinenumbers[10] 
\usepackage{nomencl} 
\usepackage{framed} 
\usepackage{multicol} 
\usepackage{etoolbox} 
\usepackage{amsmath} 
\usepackage{graphicx} 
\usepackage{tabularx} 
\newcolumntype{Y}{>{\centering\arraybackslash}X} 
\usepackage{multirow}
\usepackage{enumitem} 
\usepackage{subcaption} 
\usepackage{bm} 
\usepackage{float} 
\usepackage{physics}
\usepackage{pdflscape}
\usepackage{color}

\makenomenclature
\renewcommand*\nompreamble{\begin{multicols}{2}} 
	\renewcommand*\nompostamble{\end{multicols}}

\journal{International Journal of Hydrogen Energy}
\title{An advanced 1D physics-based model for PEM hydrogen fuel cells with enhanced overvoltage prediction}

\author[FEMTO,LIS]{Raphaël Gass \corref{mycorrespondingauthor}}
\ead{raphael.gass@univ-reunion.fr}
\cortext[mycorrespondingauthor]{Corresponding authors. \\ E-mail addresses: raphael.gass@univ-reunion.fr (R. Gass) and zhongliang.li@univ-fcomte.fr (Z. Li) \\ Website: https://gassraphael.github.io/}

\author[FEMTO]{Zhongliang Li \corref{mycorrespondingauthor}}
\ead{zhongliang.li@univ-fcomte.fr}
\author[LIS]{Rachid Outbib}
\author[FEMTO]{Samir Jemei}
\author[FEMTO,Institut]{Daniel Hissel}

\address[FEMTO]{Université de Franche-Comté, UTBM, CNRS, institut FEMTO-ST, FCLAB, Belfort, France}
\address[LIS]{Aix Marseille Univ, CNRS, LIS, Marseille, France}
\address[Institut]{Institut Universitaire de France, France}

\begin{document}


\begin{highlights}
	\item A 1D physics-based PEM fuel cell system model is built for a speed-accuracy trade-off.
	\item The model is fully described, without intermediary software, for greater flexibility.
	\item A limiting coefficient $s_{lim}$ is introduced to better model voltage losses.
	\item The model's static behavior is experimentally validated using polarization curves.
	\item A discussion on the evolution of simulated internal states is given.
\end{highlights}

\begin{frontmatter}
	\begin{abstract}
		A one-dimensional, dynamic, two-phase, isothermal model of proton exchange membrane fuel cell systems using a finite-difference approach has been developed. This model balances the simplicity of lumped-parameter models with the detailed accuracy of computational fluid dynamics models, offering precise internal state descriptions with low computational demand. The model's static behavior is validated experimentally using polarization curves. In addition, a novel physical parameter, the limit liquid water saturation coefficient ($s_{lim}$), is introduced in the overvoltage calculation, replacing the traditional limit current density coefficient ($i_{lim}$). This new parameter links the voltage drop at high current densities to the amount of liquid water present in the catalyst layers and the operating conditions of the fuel cell. Additionally, it has been observed that $s_{lim}$ is influenced at least by the gas pressure applied by the operator. This newly established link is promising for optimizing the control and thereby improving the performance of fuel cells. 
	\end{abstract}
	
\end{frontmatter}

\section*{Introduction}

To address the environmental consequences of human activities and promote sustainable development, it is imperative to reconsider our current unsustainable energy consumption practices. In this context, hydrogen-based technologies, particularly proton exchange membrane fuel cells (PEMFCs), show potential as a viable alternative to traditional oil usage. 
However, these technologies face technological obstacles that need to be overcome for large-scale commercialization. For instance, it is necessary to be able to operate PEMFCs at higher power and current densities. To achieve this, the European Union aims to reach $1.2$ $W.cm^{-2}$ @ $0.675$ $V$ by 2030 \cite{europeanunionCleanHydrogenJoint}, while Japan targets $6$ $kW.l^{-1}$ and $3.8$ $A.cm^{-2}$ for the same year \cite{jiaoDesigningNextGeneration2021}. However, during operation at high current density, PEM fuel cells are prone to experiencing water flooding and oxygen starvation. This susceptibility arises from the rapid electrochemical reactions occurring, leading to performance issues that can be detrimental. One way to manage this is to design models that provide information about the internal states of the stack, where physical sensors cannot be placed. With this information, the diagnostics of PEMFC can be improved, allowing for better dynamic control to enhance the stack performance \cite{feigaoMultiphysicDynamic1D2010,lunaNonlinearPredictiveControl2016}.

Ideally, it would be advisable to always utilize the most accurate PEMFC models that capture the 3D and dynamic characteristics of the stack. These models are considered the most precise available, although the current limits of understanding of fuel cell physics constrains their accuracy. However, these models \cite{wuMathematicalModelingTransient2009, fanCharacteristicsPEMFCOperating2017}, which rely on commercial software, demand significant computational resources and processing time, making them incompatible with embedded applications. To mitigate this computational burden, partial spatial reductions have been proposed. This involves combining, for example, a 3D model of the gas channels (GC) and gas diffusion layers (GDL) with a 1D model of the catalytic layers (CL) and membrane, forming a so-called "3D+1D" model \cite{xie3D+1DModelingApproach2020}. Similarly, "2D+1D" models have also been introduced \cite{robinDevelopmentExperimentalValidation2015, schumacher2+1DModellingPolymer2012}. Other researchers have suggested pseudo-3D ("P3D") models, which, in practice, correspond to multilayered 2D models \cite{tardyInvestigationLiquidWater2022}, or simply models exclusively in 2D \cite{mayurMultitimescaleModelingMethodology2015, baoTwodimensionalModelingPolymer2015}. Reductive assumptions have also been incorporated, such as stationary, isothermal models with a single phase for water. While these models effectively reduce computational load while maintaining precision in the stack's internal states, they still rely on commercial software and remain too time-consuming for practical use in embedded conditions. They require, for instance, several hours on a high-performance desktop computer to yield results in the case of stationary models.
On the other hand, there are highly simplified models that can run quickly on any computer. These are the lumped-parameter models. Among them, the so-called "0D" models physically represent the matter evolution but without modeling the spatial variations within each component. They provide a dynamic view of matter transport as well as a direct representation of the auxiliaries that enable stack control. The foundational work of Pukrushpan et al. \cite{pukrushpanControlOrientedModelingAnalysis2004}, whose model is accessible in open-source, has been widely disseminated. However, it is valuable to consider the spatial evolution of the stack's internal states along its thickness because matter variations are significant, and the physical phenomena occurring there are different. To achieve sufficiently precise control of PEMFCs, it seems crucial to retain at least this spatial direction.

To consider the distributed parameters along the stack thickness, 1D, "1D+0D," and "1D+1D" models have been studied. The "1D+0D"  \cite{grimmWaterManagementPEM2020} and "1D+1D" models \cite{lottinModellingOperationPolymer2009, shamardinaModelHighTemperaturePEM2012, wangQuasi2DTransientModel2018, yangModelingProtonExchange2019, yangInvestigationPerformanceHeterogeneity2020} from the literature are either fast but stationary \cite{grimmWaterManagementPEM2020, lottinModellingOperationPolymer2009, shamardinaModelHighTemperaturePEM2012} or dynamic but employ numerical solution methods that excessively slow down the model \cite{wangQuasi2DTransientModel2018, yangModelingProtonExchange2019, yangInvestigationPerformanceHeterogeneity2020}, rendering them incomplete for dynamic control design in both cases. As for the 1D models \cite{falcaoWaterTransportPEM2009, feigaoMultiphysicDynamic1D2010,falcaoWaterManagementPEMFC2016, xuRobustControlInternal2017, shaoComparisonSelfhumidificationEffect2020, xuReduceddimensionDynamicModel2021, vanderlindenProtonexchangeMembraneFuel2022}, some are also (partially) stationary \cite{falcaoWaterTransportPEM2009,feigaoMultiphysicDynamic1D2010,falcaoWaterManagementPEMFC2016}. Others incompletely represent matter transports within the MEA \cite{xuRobustControlInternal2017} or neglect to include the modeling of auxiliaries or bipolar plates \cite{feigaoMultiphysicDynamic1D2010,vanderlindenProtonexchangeMembraneFuel2022}. Finally, some models, such as these proposed by Y. Shao et al. and L. Xu et al. \cite{shaoComparisonSelfhumidificationEffect2020, xuReduceddimensionDynamicModel2021}, are the ones closest to the set objectives: they are fast, dynamic, biphasic, account for the balance of plant and provide sufficiently precise information on all internal states of the stack. However, it is worth noting that their proposed liquid water modeling necessitates the introduction of simplifying assumptions, such as quasi-static equilibrium or an infinite evaporation rate. It is essential to alleviate these assumptions by incorporating insights from alternative 1D models \cite{vanderlindenProtonexchangeMembraneFuel2022} that consider liquid water without resorting to such reductive assumptions. This ensures the credibility of the model predictions.

One objective of the present work is to overcome the drawbacks of the above modelings by developing a comprehensive model of the PEM fuel cell system that eliminates the previous simplifying assumptions regarding the evolution of liquid water, while still maintaining its speed qualities. This model is 1D, dynamic, biphasic, and isothermal. In the developed model, certain involved equations are revised or improved, incorporating findings from recent research and extending upon the authors' prior work \cite{gassCriticalReviewProton2024}. Some original equations have been added and discussed concerning auxiliary variables and voltage calculation to make the model more comprehensive and realistic. In particular, a novel coefficient, named the limit liquid water saturation coefficient ($s_{lim}$), is introduced to better model the voltage drop at high current densities, establishing a connection between this current density limit and the internal states as well as operating conditions of the cell. Ultimately, this open-source model has been designed to be adopted and extended by other researchers to expedite research in this field. In particular, the coupling of physics-based models like this one with machine learning-based models appears highly promising for producing even faster models while maintaining a very high level of accuracy \cite{wangIntegrationMultiphysicsMachine2023,wangLongShorttermMemory2023,karniadakisPhysicsinformedMachineLearning2021}.
\section{Modeling matter flow in a PEM cell}
The model developed in this study is oriented to real-time diagnosis and control purposes. It is therefore needed to take into account both execution speed and accuracy. For instance, regarding the mass transfer process, the model is expected to predict the next tens to hundreds seconds within a few seconds. This enables the controllers to perform multiple model-base predictions within a single control period so that a model predictive control paradigm can be deployed. However, these predictions must also be sufficiently accurate to support the model based diagnosis and control to avoid unintentionally putting the stack in a faulty state or a highly degraded condition, as well as preventing hydrogen waste.

To fulfill these requirements, a one-dimensional (1D) model has been proposed. To achieve efficient gas and water management-related control, real-time access to the dynamically varying spatial distribution of internal states within the fuel stack is necessary. These states encompass the concentrations of reactants and products, the proportion of liquid or dissolved water in the membrane, and the flow of matter throughout the stack. These variables primarily evolve in the thickness direction of the stack, which is why a 1D model was selected. Furthermore, the condensation of water vapor within the stack is important to consider as flooding must be closely monitored. As a result, the model accounts for two states of water molecules: vapor and liquid, making it a two-phase model. Lastly, it is important to note that the model assumes isothermal conditions and considers that all cell exhibit identical behavior throughout the entire stack. These significant assumptions were made to simplify the complexity of developing the model and are expected to be eliminated in future model versions.  

For the model resolution, a finite-difference method is employed to discretize the partial differential equations governed model and transform it into an ordinary differential equations (ODE) governed one. The number and positions of nodes were set appropriately to simplify the model resolution to the utmost extent without losing accuracy. An adaptable numerical method is then applied to solve the transformed ODE.

In the sequel, the finite-difference method, the numerical solution, and the transformed model are presented successively. The balance of plant modeling is discussed in section \ref{sec:auxiliary_modeling}.

\subsection{Finite-difference model and its numerical solution}
\subsubsection{Finite-difference modeling method}
\label{subsubsec:nodal_model}
Finite-difference modeling involves dividing a system into discrete nodes, with each node representing a specific volume within the system. Within each region, all quantities are assumed to be homogeneous. The value at the center of each volume is then extrapolated to the entire one. Consequently, each node is positioned at the center of its respective region. Therefore, by decreasing the size of the volumes, the simplifying assumption becomes less significant, resulting in a more accurate model.

Within a PEM single cell, there are seven distinct zones. The anode consists of a GDL and a CL. It is in contact with a gas channel (GC) on one side and a membrane on the other side. The configuration is similar on the cathode side, and a single membrane separates the anode from the cathode within the same cell. Each of these zones is composed of different materials or experiences the flow of different molecules. To accurately represent these structures and the matter flow within them, each zone must be assigned a separate node at minimum since each node homogenizes the quantities present within it. Therefore, a minimum of seven nodes is required, corresponding to the seven zones under consideration.

Then, it is also necessary to include an additional node at each GDL, specifically at the boundary with the bipolar plate. These additional nodes are required to account for the material discontinuity between the GDL and the GC, which results in sorption flows between them. Including these nodes accurately captures the sorption flows and ensures the model properly represents this phenomenon.

Furthermore, due to the difference in thickness between the GDL and the CL, it is not enough to only use 9 nodes. Indeed, for the sake of numerical stability, it is advisable to have distances between the nodes of the discretization scheme that are of the same order of magnitude. Ideally, each GDL should have a number of nodes, denoted as $n_{gdl}$, equal to $\lfloor \frac{H_{gdl}}{H_{cl}} \rfloor$. However, this results in a large number of nodes within the cell, with $n_{gdl}$ generally exceeding 20. Given the number of variables interacting in the GDL, this has a significant computational time cost. In line with the compromise approach of this study, the authors thus propose to take $n_{gdl} = \lfloor \frac{H_{gdl}}{2H_{cl}} \rfloor$.

\begin{figure}[H]
	\centering
	\includegraphics[width=16.5cm]{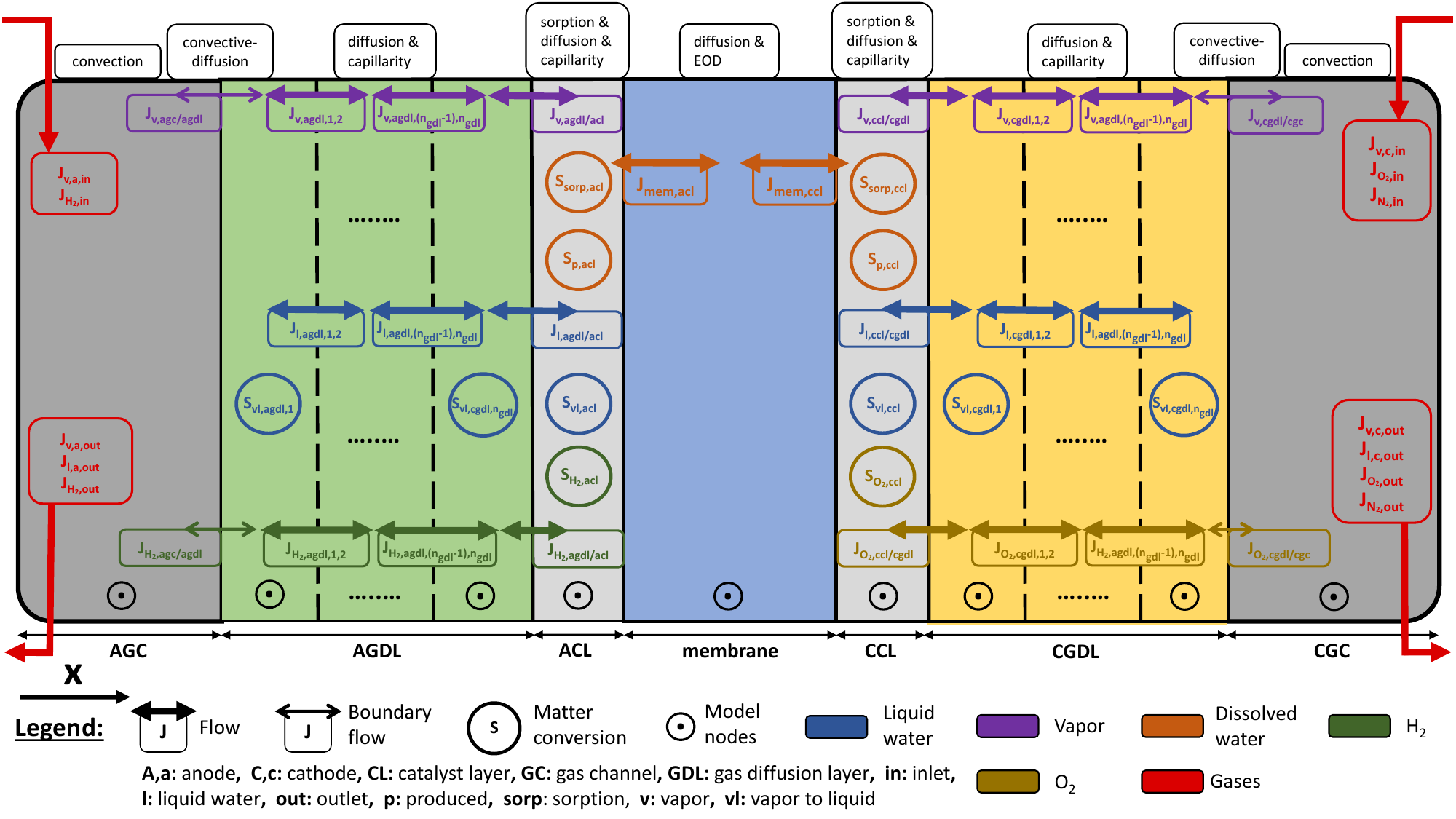}
	\caption{1D modeling of matter transport phenomena in a PEM single cell divided into several nodes}
	\label{fig:nodal_model}
\end{figure}

Finally, figure \ref{fig:nodal_model} was generated to illustrate both the overall flows and matter conversions, including their notations, and the placement of model nodes within a PEM single cell. 

\subsubsection{Numerical solution method}

To solve the finite-difference model, the '$BDF$' (Backward Differentiation Formula) method, available in the '$\verb|solve_ivp|$' function of Python's scipy.integrate module, has been utilized \cite{scipyScipyIntegrateSolve_ivp}. This method offers several advantages. 

Firstly, it is an implicit method that guarantees the convergence of results, which is particularly valuable for this model as it involves a stiff problem with high sensitivity to parameters. Indeed, the various physical phenomena in the fuel stack are interconnected. For instance, the consumption of hydrogen leads to the production of dissolved water, which subsequently influences the amount of water vapor or liquid water present. Furthermore, matters evolve at different timescales in the whole fuel cell system. Gases, for example, move much faster compared to liquid water in the stack. This complexity gives rise to a stiff problem that necessitates meticulous numerical solving techniques.

Secondly, this '$BDF$' method employs a non-constant step size, automatically identifying regions of significant changes that require smaller time steps, as well as regions with more gradual changes where larger time steps can be used. This results in a significant reduction in computation time.

Finally, it is important to remember that only methods that can handle stiff problems can be used to solve the proposed model, which excludes most explicit methods.

\subsection{The flows and differential equations at stake}

\subsubsection{Working hypotheses}
The assumptions made for the model are listed as follows. The assumptions that were used to develop the mathematical expressions of the flows and differential equations are not mentioned here and are present in the authors' previous work \cite{gassCriticalReviewProton2024}.

\textbf{Overall}
\begin{itemize}[noitemsep]
	\item The cells in the concerned stack are identical, in terms of parameters and operating conditions.
	\item The stack temperature is assumed to be constant and uniform. 
	\item All gas species behave ideally.
	\item The effect of gravity is ignored.
	\item Nitrogen concentration is deemed homogeneous across both the cathode and the cathode bipolar plate, with no spatial variation being considered.
\end{itemize}

\textbf{In the membrane}
\begin{itemize}[noitemsep]
	\item The membrane is considered to be perfectly impermeable to electrons, neglecting the internal short circuit.
	\item The water generated in the triple point region of the cathode is assumed to be produced in dissolved form in the membrane \cite{jiaoWaterTransportPolymer2011}.
	\item The flow of water through the membrane to a catalytic layer is assumed to be a flow of dissolved water which becomes vapor water \cite{geAbsorptionDesorptionTransport2005}.
	\item Since the catalytic layer is very thin compared to the other layers, it is considered that the $\lambda$ value of the electrolyte present in the CL is instantly the same as at the membrane boundary \cite{xuReduceddimensionDynamicModel2021} : 
	\[\lambda_{acl} = \lim_{x \to acl} \lambda_{mem} \text{ and } 
	\lambda_{ccl} = \lim_{x \to ccl} \lambda_{mem} 
	\]
\end{itemize}

\textbf{In the GCs}
\begin{itemize}[noitemsep]
	\item Liquid water is considered nonexistent in the GC, and a Dirichlet boundary condition \cite{chengHeritageEarlyHistory2005} is imposed at the GDL/GC interface, setting the liquid water saturation variable \texttt{s} to zero.
	\item All gases move at the same speed through the GC.
	\item Water phase change is ignored in the GC.
	\item All concentrations are uniform in the GC.
\end{itemize}

\subsubsection{Adaptation of mathematical expressions to the finite-difference model}
\label{subsec:adaptation_expressions_model}
To solve the system of differential equations that describes the matter transports in the stack \cite{gassCriticalReviewProton2024}, certain simplifications have been applied to tailor the mathematical expressions to the proposed finite-difference model.

Firstly, the spatial gradients $\grad$ have been approximated using a partial spatial derivative through the thickness of the cell, denoted as $\frac{\partial}{\partial x} \bm{\imath}$, where $\bm{\imath}$ is a unit vector pointing from the anode to the cathode direction. This simplification is valid because the main circulation of matter occurs along this spatial direction, $x$. The notation $\partial$ is retained to indicate that the quantities involved are dependent on other variables, such as time $t$. Subsequently, this partial derivative $\frac{\partial}{\partial x}$ is replaced by a finite difference between two nodes. These successive simplifications are illustrated in Equation \eqref{eq:example_simplifications}, which describes the diffusion of water vapor in the anode:
\begin{equation}
	\bm{J_{dif}} = - D_{v}^{eff} \bm{\nabla} C_v \approx - D_{v}^{eff} \frac{\partial C_v}{\partial x} \bm{\imath} \approx - 2 D_{v}^{eff} \frac{C_{v,acl} - C_{v,agdl,n_{gdl}}}{H_{gdl}/n_{gdl} + H_{cl}}  \bm{\imath}
	\label{eq:example_simplifications}
\end{equation}
where $C_{agdl,n_{gdl}}$ is the vapor concentration at the $n_{gdl}$-th node of the AGDL.

Furthermore, in the calculation of flow between two nodes, many parameters or variables need to be averaged. For instance, in the case of water vapor diffusion mentioned earlier, the effective diffusion coefficient $D_{v}^{eff}$ is dependent on several factors, including liquid water saturation $s$, porosity $\varepsilon$, pressure $P$, and temperature $T$: $D_{v}^{eff}(s,\varepsilon,P,T)$. These four quantities, among others, vary spatially. However, when studying the flow between two nodes, it is necessary to assign a single symmetric value for $D_{v}^{eff}$. The proposed approach is to average the variables and parameters of two consecutive nodes. Thus, secondary variables and parameters are introduced, as seen in \eqref{eq:example_adapted_expression} with $\texttt{s}_{agdl,acl}$, $\varepsilon_{agdl,acl}$, $P_{agdl,acl}$ and $T_{agdl,acl}$. In this study, the spatial variation of temperature is implied, although the model assumes an isothermal condition. This is made to facilitate the future implementation of heat transfers.

\begin{equation}
	\begin{cases}
		\bm{J_{dif}} = - D_{v}^{eff}(\texttt{s},\varepsilon,P,T) \bm{\nabla} C_v \approx - 2 D_{v}^{eff}(\texttt{s}_{agdl,acl},\varepsilon_{agdl,acl},P_{agdl,acl},T_{agdl,acl}) \frac{C_{v,acl} - C_{v,agdl,n_{gdl}}}{H_{gdl}/n_{gdl} + H_{cl}} \bm{\imath} \\
		
		\texttt{s}_{agdl,acl} = \frac{\texttt{s}_{agdl,n_{gdl}} + \texttt{s}_{acl}}{2}, 
		\varepsilon_{agdl,acl} = \frac{\varepsilon_{agdl} + \varepsilon_{acl}}{2}, \\
		P_{agdl,acl} = \frac{P_{agdl,n_{gdl}} + P_{acl}}{2},
		T_{agdl,acl} = \frac{T_{agdl,n_{gdl}} + T_{acl}}{2}
	\end{cases}
	\label{eq:example_adapted_expression}
\end{equation}

\subsubsection{Expression of the physical phenomena involved}

After incorporating the previously discussed modifications, the differential equations and matter transport expressions outlined in the authors' earlier work \cite{gassCriticalReviewProton2024} can be represented as shown in tables \ref{table:synthesis_flows_equadif_1} and \ref{table:synthesis_flows_equadif_2}. It should be noted that here the parameter $L_{gc}$ represents the cumulative length of the gas channel, which is the total distance traveled by the gases as they circulate through the bipolar plates. Additionally, the flow coefficients that are functions of internal states have been adjusted for this model and are provided in table \ref{table:synthesis_flow_coef}. Finally, general parameters for modeling the cell are furnished in table \ref{table:synthesis_cell_parameters}, while the cell's specific parameters contingent upon the cell type should be identified independently. This will be discussed in Section \ref{sec:validation_model}.

\begin{landscape}
	\begin{table}[H]
		\centering
		\small
		\linespread{1.5}
		\begin{tabularx}{\linewidth}{|Y|YY|} \hline
			
			\bf{Dynamical models} &
			\multicolumn{2}{>{\hsize=\dimexpr2\hsize+2\tabcolsep+\arrayrulewidth\relax}Y|}{
				\bf{Matter flow expressions}} \\ \hline \hline

			\multicolumn{3}{|c|}{\bf{Dissolved water in the membrane}} \\ \hline

			\multirow{2}{*}{
				$\frac{\rho_{mem} \varepsilon_{mc}}{M_{eq}} \frac{d \lambda_{acl}}{d t} = - \frac{J_{\lambda,mem,acl}}{H_{cl}} + S_{sorp,acl} + S_{p,acl}$ \label{subeq:water_content_dynamic_balance_lambda4}} &
			\multicolumn{2}{>{\hsize=\dimexpr2\hsize+2\tabcolsep+\arrayrulewidth\relax}Y|}{
				$S_{\text{p},acl} = 2 k_{O_{2}}\left( \lambda_{mem}, T_{fc} \right) \frac{R T_{fc}}{H_{cl} H_{mem}} C_{O_{2},ccl} $ \label{eq:S_prod_acl}} \\
			
			&
			\multicolumn{2}{>{\hsize=\dimexpr2\hsize+2\tabcolsep+\arrayrulewidth\relax}Y|}{
				$S_{sorp,acl} = \gamma_{sorp}(\lambda_{acl},T_{fc})  \frac{\rho_{\text{mem}}}{M_{\text{eq}}} \left[ \lambda_{\text{eq}}(C_{v,acl},\texttt{s}_{acl},T_{fc})  - \lambda_{acl} \right]$ \label{eq:j_sorp_Ge_a}} \\
			
			\multirow{2}{*}{
				$\frac{\rho_{mem}}{M_{eq}} \frac{d \lambda_{mem}}{d t} = \frac{J_{\lambda,mem,acl} - J_{\lambda,mem,ccl}}{H_{mem}}$ \label{subeq:water_content_dynamic_balance_lambda_mem}} &
			\multicolumn{2}{>{\hsize=\dimexpr2\hsize+2\tabcolsep+\arrayrulewidth\relax}Y|}{
				$J_{\lambda,mem,acl} = \frac{2.5}{22} \frac{i_{fc}}{F} \lambda_{acl,mem} - \frac{2\rho_{mem}}{M_{eq}} D(\lambda_{acl,mem}) \frac{\lambda_{mem}-\lambda_{acl}}{H_{mem}+H_{cl}}
				\label{eq:water_flow_membrane_acl}$}  \\
			
			&
			\multicolumn{2}{>{\hsize=\dimexpr2\hsize+2\tabcolsep+\arrayrulewidth\relax}Y|}{
				$J_{\lambda,mem,ccl} = \frac{2.5}{22} \frac{i_{fc}}{F} \lambda_{mem,ccl} - \frac{2\rho_{mem}}{M_{eq}} D(\lambda_{mem,ccl}) \frac{\lambda_{ccl}-\lambda_{mem}}{H_{mem}+H_{cl}}$
				\label{eq:water_flow_membrane_ccl}}  \\
			
			\multirow{2}{*}{
				$\frac{\rho_{mem} \varepsilon_{mc}}{M_{eq}} \frac{d \lambda_{ccl}}{d t} = \frac{J_{\lambda,mem,ccl}}{H_{cl}} + S_{sorp,ccl} + S_{p,ccl}$ \label{subeq:water_content_dynamic_balance_lambda_ccl}} &	
			\multicolumn{2}{>{\hsize=\dimexpr2\hsize+2\tabcolsep+\arrayrulewidth\relax}Y|}{
				$S_{sorp,ccl} = \gamma_{sorp}(\lambda_{ccl},T_{fc})  \frac{\rho_{\text{mem}}}{M_{\text{eq}}} \left[ \lambda_{\text{eq}}(C_{v,ccl},\texttt{s}_{ccl},T_{fc}) - \lambda_{ccl} \right]$ \label{eq:j_sorp_Ge_c}} \\
			
			&
			\multicolumn{2}{>{\hsize=\dimexpr2\hsize+2\tabcolsep+\arrayrulewidth\relax}Y|}{
				$S_{\text{p},ccl} = \frac{i_{fc}}{2 F H_{cl}} + k_{H_{2}}\left( \lambda_{mem}, T_{fc} \right) \frac{R T}{H_{cl} H_{mem}} C_{H_{2},acl}$ \label{eq:S_prod_ccl}} \\ \hline

			\multicolumn{3}{|c|}{\bf{Liquid water in the GDL and the CL}} \\ \hline
			
			\small{
				$\forall i \in \llbracket 2, n_{\text{gdl}}-1 \rrbracket :$
				$\rho_{H_{2}O} \varepsilon_{gdl} \frac{d \texttt{s}_{agdl,i}}{d t} = \frac{J_{l,agdl,\left( i-1 \right),i} - J_{l,agdl,i,\left( i+1 \right)}}{H_{gdl}/n_{gdl}} + M_{H_{2}O} S_{vl,agdl,i} $} &
			\multicolumn{2}{>{\hsize=\dimexpr2\hsize+2\tabcolsep+\arrayrulewidth\relax}Y|}{	
				\small{ 
					$\forall i \in \llbracket 1, n_{\text{gdl}}-1 \rrbracket :$
					$J_{l,agdl,i,\left( i+1 \right)} = \sigma(T_{fc}) \frac{K_{0}(\varepsilon_{gdl})}{\nu_{l}} \cos \left( \theta_{c,gdl} \right) \sqrt{\frac{\varepsilon_{gdl}}{K_{0}(\varepsilon_{gdl})}} \texttt{s}_{agdl,i,\left( i+1 \right)}^{\texttt{e}} \left[ 1.417 - 4.24\texttt{s}_{agdl,i,\left( i+1 \right)} + 3.789\texttt{s}_{agdl,i,\left( i+1 \right)}^{2} \right] \frac{\texttt{s}_{agdl,\left(i+1\right)} - \texttt{s}_{agdl,i}}{H_{gdl}/n_{gdl}}$ \label{eq:Jl,agdl,agdl,i}}} \\

			\scriptsize{
				$\rho_{H_{2}O} \varepsilon_{gdl} \frac{d \texttt{s}_{agdl,n_{gdl}}}{d t} = \frac{J_{l,agdl,\left( n_{gdl}-1 \right),n_{gdl}} - J_{l,agdl,acl}}{H_{gdl}/n_{gdl}} + M_{H_{2}O} S_{vl,agdl,n_{gdl}} $} &
			\multicolumn{2}{>{\hsize=\dimexpr2\hsize+2\tabcolsep+\arrayrulewidth\relax}Y|}{
				\small{
					$J_{l,agdl,acl} = 2 \sigma(T_{fc}) \frac{K_{0}(\varepsilon_{gdl,cl})}{\nu_{l}} \cos \left( \theta_{c,gdl,cl} \right) \sqrt{\frac{\varepsilon_{gdl,cl}}{K_{0}(\varepsilon_{gdl,cl})}} \texttt{s}_{agdl,acl}^{\texttt{e}} \left[ 1.417 - 4.24\texttt{s}_{agdl,acl} + 3.789\texttt{s}_{agdl,acl}^{2} \right] \frac{\texttt{s}_{acl} - \texttt{s}_{agdl,n_{gdl}}}{H_{gdl}/n_{gdl} + H{cl}}$ \label{eq:Jl_34}}}  \\

			\small{
				$\rho_{H_{2}O} \varepsilon_{cl} \frac{d \texttt{s}_{acl}}{d t} = \frac{J_{l,agdl,acl}}{H_{cl}} + M_{H_{2}O} S_{vl,acl} $} &
			\multicolumn{2}{>{\hsize=\dimexpr2\hsize+2\tabcolsep+\arrayrulewidth\relax}Y|}{
				\small{
					$J_{l,ccl,cgdl} = 2 \sigma(T_{fc}) \frac{K_{0}(\varepsilon_{gdl,cl})}{\nu_{l}} \cos \left( \theta_{c,gdl,cl} \right) \sqrt{\frac{\varepsilon_{gdl,cl}}{K_{0}(\varepsilon_{gdl,cl})}} \texttt{s}_{ccl,cgdl}^{\texttt{e}} \left[ 1.417 - 4.24\texttt{s}_{ccl,cgdl} + 3.789\texttt{s}_{ccl,cgdl}^{2} \right] \frac{\texttt{s}_{cgdl,1} - \texttt{s}_{ccl}}{H_{gdl}/n_{gdl} + H{cl}}$ \label{eq:Jl_67}}} \\

			\small{
				$\rho_{H_{2}O} \varepsilon_{cl} \frac{d \texttt{s}_{ccl}}{d t} = \frac{-J_{l,ccl,cgdl}}{H_{cl}} + M_{H_{2}O} S_{vl,ccl} $} &
			\multicolumn{2}{>{\hsize=\dimexpr2\hsize+2\tabcolsep+\arrayrulewidth\relax}Y|}{	
				\small{
					$J_{l,cgdl,i,\left( i+1 \right)} = \sigma(T_{fc}) \frac{K_{0}(\varepsilon_{gdl})}{\nu_{l}} \cos \left( \theta_{c,gdl} \right) \sqrt{\frac{\varepsilon_{gdl}}{K_{0}(\varepsilon_{gdl})}} \texttt{s}_{cgdl,i,\left( i+1 \right)}^{\texttt{e}} \left[ 1.417 - 4.24\texttt{s}_{cgdl,i,\left( i+1 \right)} + 3.789\texttt{s}_{cgdl,i,\left( i+1 \right)}^{2} \right] \frac{\texttt{s}_{cgdl,\left(i+1\right)} - \texttt{s}_{cgdl,i}}{H_{gdl}/n_{gdl}}$ \label{eq:Jl,cgdl,cgdl,i}}} \\

			\small{
				$\rho_{H_{2}O} \varepsilon_{gdl} \frac{d \texttt{s}_{cgdl,1}}{d t} = \frac{J_{l,ccl,cgdl} - J_{l,cgdl,1,2}}{H_{gdl}/n_{gdl}} + M_{H_{2}O} S_{vl,cgdl,1} $} & 
			\multicolumn{2}{>{\hsize=\dimexpr2\hsize+2\tabcolsep+\arrayrulewidth\relax}Y|}{
				\multirow{3}{*}{
					\small{
						$S_{vl} = 
						\begin{cases}
							\gamma_{\text{cond}} \varepsilon \left( 1 - \texttt{s} \right) x_{v} \left( C_v - C_{\text {v,sat}} \right), & \text{if $C_v > C_{\text{v,sat}}$} \\
							-\gamma_{\text{evap}} \varepsilon \texttt{s} \frac{\rho_{H_{2}O}}{M_{H_{2}O}} R T_{fc} \left( C_{\text{v,sat}} - C_v \right), &\text{if $C_v \leq C_{\text{v,sat}}$}
						\end{cases}$
						\label{eq:phase_change_rate}}}} \\

			\small{
				$\rho_{H_{2}O} \varepsilon_{gdl} \frac{d \texttt{s}_{cgdl,i}}{d t} = \frac{J_{l,cgdl,\left( i-1 \right),i} - J_{l,cgdl,i,\left( i+1 \right)}}{H_{gdl}/n_{gdl}} + M_{H_{2}O} S_{vl,cgdl,i} $} & & \\

			\small{
				Boundary conditions: $\texttt{s}_{agdl,1} = 0$, $\texttt{s}_{cgdl,n_{gdl}} = 0$} & & \\ \hline

			\multicolumn{3}{|c|}{\bf{Vapor in the GC}} \\ \hline
			
			\multirow{2.5}{*}{
				$\frac{\mathrm{d}C_{v,agc}}{\mathrm{dt}} = \frac{J_{v,a,\text{in}} - J_{v,a,\text{out}}}{L_{gc}} - \frac{J_{v,agc,agdl}}{H_{gc}} $} &
			\multicolumn{2}{>{\hsize=\dimexpr2\hsize+2\tabcolsep+\arrayrulewidth\relax}Y|}{
				$ J_{v,a,\text{in}} = \frac{\Phi_{asm} P_{sat}\left( T_{fc} \right)}{P_{asm}} \frac{W_{asm,out}}{H_{gc} W_{gc} M_{asm}} $} \\ 
			
			& 
			\multicolumn{2}{>{\hsize=\dimexpr2\hsize+2\tabcolsep+\arrayrulewidth\relax}Y|}{
				$ J_{v,a,\text{out}} = \frac{\Phi_{agc} P_{sat}\left( T_{fc} \right)}{P_{agc}} \frac{W_{aem,in}}{H_{gc} W_{gc} M_{agc}} $}  \\

			\multirow{2}{*}{
				$\frac{\mathrm{d}C_{v,cgc}}{\mathrm{dt}} = \frac{J_{v,c,\text{in}} - J_{v,c,\text{out}}}{L_{gc}} + \frac{J_{v,cgdl,cgc}}{H_{gc}} $} &
			\multicolumn{2}{>{\hsize=\dimexpr2\hsize+2\tabcolsep+\arrayrulewidth\relax}Y|}{
				$ J_{v,c,\text{in}} = \frac{\Phi_{csm} P_{sat}\left( T_{fc} \right)}{P_{csm}} \frac{W_{csm,out}}{H_{gc} W_{gc} M_{csm}} $} \\
			
			&
			\multicolumn{2}{>{\hsize=\dimexpr2\hsize+2\tabcolsep+\arrayrulewidth\relax}Y|}{
				$ J_{v,c,\text{out}} = \frac{\Phi_{cgc} P_{sat}\left( T_{fc} \right)}{P_{cgc}} \frac{W_{cem,in}}{H_{gc} W_{gc} M_{cgc}} $} \\ \hline

			\multicolumn{3}{|c|}{\bf{Hydrogen and oxygen in the GC}} \\ \hline
			
			\multirow{2.5}{*}{
				$\frac{\mathrm{d}C_{H_{2},agc}}{\mathrm{dt}} = \frac{J_{H_{2},\text{in}} - J_{H_{2},\text{out}}}{L_{gc}} - \frac{J_{H_{2},agc,agdl}}{H_{gc}} $} &
			\multicolumn{2}{>{\hsize=\dimexpr2\hsize+2\tabcolsep+\arrayrulewidth\relax}Y|}{
				$ J_{H_{2},\text{in}} = \frac{P_{asm} - \Phi_{asm}P_{sat}\left(T_{fc}\right)}{P_{asm}} \frac{W_{asm,out}}{H_{gc} W_{gc} M_{asm}} $} \\  
			
			& 
			\multicolumn{2}{>{\hsize=\dimexpr2\hsize+2\tabcolsep+\arrayrulewidth\relax}Y|}{
				$ J_{H_{2},\text{out}} = \frac{P_{agc} - \Phi_{agc}P_{sat}\left(T_{fc}\right)}{P_{agc}} \frac{W_{aem,in}}{H_{gc} W_{gc} M_{agc}} $}  \\ 
			
			\multirow{2.5}{*}{
				$\frac{\mathrm{d}C_{O_{2},cgc}}{\mathrm{dt}} = \frac{J_{O_{2},\text{in}} - J_{O_{2},\text{out}}}{L_{gc}} + \frac{J_{O_{2},cgdl,cgc}}{H_{gc}}$} &
			\multicolumn{2}{>{\hsize=\dimexpr2\hsize+2\tabcolsep+\arrayrulewidth\relax}Y|}{
				$ J_{O_{2},\text{in}} = y_{O_{2},ext} \frac{P_{csm} - \Phi_{csm}P_{sat}\left(T_{fc}\right)}{P_{csm}} \frac{W_{csm,out}}{H_{gc} W_{gc} M_{csm}} $} \\
			
			&
			\multicolumn{2}{>{\hsize=\dimexpr2\hsize+2\tabcolsep+\arrayrulewidth\relax}Y|}{
				$ J_{O_{2},\text{out}} = y_{O_{2},cgc} \frac{P_{cgc} - \Phi_{cgc}P_{sat}\left(T_{fc}\right)}{P_{cgc}} \frac{W_{cem,in}}{H_{gc} W_{gc} M_{cgc}} $} \\ \hline

			\multicolumn{3}{|c|}{\bf{Nitrogen}} \\ \hline
			
			\multirow{2.5}{*}{
				$\frac{\mathrm{d}C_{N_{2}}}{\mathrm{dt}} = \dfrac{J_{N_{2},\text{in}} - J_{N_{2},\text{out}}}{L_{gc}} $} 
			& \multicolumn{2}{>{\hsize=\dimexpr2\hsize+2\tabcolsep+\arrayrulewidth\relax}Y|}{
				$ J_{N_{2},\text{in}} = \left( 1 - y_{O_{2},ext} \right) \frac{P_{csm} - \Phi_{csm}P_{sat}\left(T_{fc}\right)}{P_{csm}} \frac{W_{csm,out}}{H_{gc} W_{gc} M_{csm}} $}  \\ 
			
			& \multicolumn{2}{>{\hsize=\dimexpr2\hsize+2\tabcolsep+\arrayrulewidth\relax}Y|}{
				$ J_{N_{2},\text{out}} = \left( 1 - y_{O_{2},cgc} \right) \frac{P_{cgc} - \Phi_{cgc}P_{sat}\left(T_{fc}\right)}{P_{cgc}} \frac{W_{cem,in}}{H_{gc} W_{gc} M_{cgc}} $} \\ \hline

		\end{tabularx}
		\caption{Synthesis of the differential equations and the associated matter transport expressions in the stack \cite{gassCriticalReviewProton2024} (1/2)}
		\label{table:synthesis_flows_equadif_1}
	\end{table}		
\end{landscape}

\begin{landscape}
	\begin{table}[H]
		\centering
		\linespread{1.5}
		\begin{tabularx}{\linewidth}{|Y|Y|} \hline		
			
			\bf{Dynamical models} & \bf{Matter flow expressions} \\ \hline \hline

			\multicolumn{2}{|c|}{\bf{Vapor in the GDL and the CL}} \\ \hline
			
			$\forall i \in \llbracket 2, n_{\text{gdl}}-1 \rrbracket :$ &
			$\forall i \in \llbracket 1, n_{\text{gdl}}-1 \rrbracket :$  \\

			$\varepsilon_{gdl} \left[ 1-\texttt{s}_{agdl,1} \right] \frac{d C_{v,agdl,1}}{d t} = \frac{J_{v,agc,agdl}-J_{v,agdl,1,2}}{H_{gdl}/n_{gdl}} - S_{vl,agdl,1} $ &
			$J_{v,agc,agdl} = h_{a}(P_{agc,agdl},T_{fc}) \left[ C_{v,agc} - C_{v,agdl,1} \right] $ \label{eq:Jv_agc,agdl}  \\
			
			$\varepsilon_{gdl} \left[ 1-\texttt{s}_{agdl,i} \right] \frac{d C_{v,agdl,i}}{d t} = \frac{J_{v,agdl,\left( i-1 \right),i}-J_{v,agdl,i\left( i+1 \right)}}{H_{gdl}/n_{gdl}} - S_{vl,agdl,i} $ &
			\multirow{3}{*}{
				$J_{v,agdl,i,\left( i+1 \right)} = - D_{a,eff}(\texttt{s}_{agdl,i,\left( i+1 \right)},\varepsilon_{gdl},P_{agdl,i,\left( i+1 \right)},T_{fc}) \frac{C_{v,agdl,\left( i+1 \right)}-C_{v,agdl,i}}{H_{gdl}/n_{gdl}}   $ \label{eq:Jv_23}}  \\ 
			
			$\varepsilon_{gdl} \left[ 1-\texttt{s}_{agdl,n_{gdl}} \right] \frac{d C_{v,agdl,n_{gdl}}}{d t} = \frac{J_{v,agdl,\left( n_{gdl}-1 \right),n_{gdl}}-J_{v,agdl,acl}}{H_{gdl}/n_{gdl}} - S_{vl,agdl,n_{gdl}} $ & \\
			
			$\varepsilon_{cl} \left[ 1-\texttt{s}_{acl} \right] \frac{d C_{v,acl}}{d t} = \frac{J_{v,agdl,acl}}{H_{cl}} - S_{sorp,acl} - S_{vl,acl} $ &
			$J_{v,agdl,acl} = - 2D_{a,eff}(\texttt{s}_{agdl,acl},\varepsilon_{agdl,acl},P_{agdl,acl},T_{fc}) \frac{C_{v,acl}-C_{v,agdl,n_{agdl}}}{H_{gdl}/n_{gdl} + H_{cl}}   $ \label{eq:Jv_agdl_acl}  \\ 
			
			$\varepsilon_{cl} \left[ 1-\texttt{s}_{ccl} \right] \frac{d C_{v,ccl}}{d t} = -\frac{J_{v,ccl,cgdl}}{H_{cl}} - S_{sorp,ccl} - S_{vl,ccl} $ &
			$J_{v,ccl,cgdl} = - 2D_{c,eff}(\texttt{s}_{ccl,cgdl},\varepsilon_{ccl,cgdl},P_{ccl,cgdl},T_{fc}) \frac{C_{v,cgdl,1}-C_{v,ccl}}{H_{gdl}/n_{gdl} + H_{cl}}   $ \label{eq:Jv_ccl_cgdl} \\ 
			
			$\varepsilon_{gdl} \left[ 1-\texttt{s}_{cgdl,1} \right] \frac{d C_{v,cgdl,1}}{d t} = \frac{J_{v,ccl,cgdl}-J_{v,cgdl,1,2}}{H_{gdl}/n_{gdl}} - S_{vl,cgdl,1} $ & 
			\multirow{3}{*}{
				$J_{v,cgdl,i,\left( i+1 \right)} = - D_{c,eff}(\texttt{s}_{cgdl,i,\left( i+1 \right)},\varepsilon_{gdl},P_{cgdl,i,\left( i+1 \right)},T_{fc}) \frac{C_{v,cgdl,\left( i+1 \right)}-C_{v,cgdl,i}}{H_{gdl}/n_{gdl}}   $ \label{eq:Jv_cgdl_i_i+1}} \\
			
			$\varepsilon_{gdl} \left[ 1-\texttt{s}_{cgdl,i} \right] \frac{d C_{v,cgdl,i}}{d t} = \frac{J_{v,cgdl,\left( i-1 \right),i}-J_{v,cgdl,i,\left( i+1 \right)}}{H_{gdl}/n_{gdl}} - S_{vl,cgdl,i} $ & \\ 
			
			$\varepsilon_{gdl} \left[ 1-\texttt{s}_{cgdl,n_{gdl}} \right] \frac{d C_{v,cgdl,n_{gdl}}}{d t} = \frac{J_{v,cgdl,\left( n_{gdl}-1 \right),n_{gdl}}-J_{v,cgdl,cgc}}{H_{gdl}/n_{gdl}} - S_{vl,cgdl,n_{gdl}} $ & 
			$J_{v,cgdl,cgc} = h_{c}(P_{cgdl,cgc},T_{fc}) \left[ C_{v,cgdl,n_{cgdl}} - C_{v,cgc} \right] $ \label{eq:Jv_cgdl_cgc} \\ \hline

			\multicolumn{2}{|c|}{\bf{Hydrogen in the GDL and the CL}} \\ \hline
			
			$\forall i \in \llbracket 2, n_{\text{gdl}}-1 \rrbracket :$ &
			$\forall i \in \llbracket 1, n_{\text{gdl}}-1 \rrbracket :$  \\
			
			$\varepsilon_{gdl} \left[ 1-\texttt{s}_{agdl,1} \right] \frac{d C_{H_{2},agdl,1}}{d t} = \frac{J_{H_{2},agc,agdl}-J_{H_{2},agdl,1,2}}{H_{gdl}/n_{gdl}} $  &
			$J_{H_{2},agc,agdl} = h_{a}(P_{agc,agdl},T_{fc}) \left[ C_{H_{2},agc} - C_{H_{2},agdl,1} \right] $ \label{eq:J_H2_agc_agdl} \\
			
			$\varepsilon_{gdl} \left[ 1-\texttt{s}_{agdl,i} \right] \frac{d C_{H_{2},agdl,i}}{d t} = \frac{J_{H_{2},agdl,\left( i-1 \right),i}-J_{H_{2},agdl,i,\left( i+1 \right)}}{H_{gdl}/n_{gdl}} $ &
			$J_{H_{2},agdl,i,\left( i+1 \right)} = - D_{a,eff}(\texttt{s}_{agdl,i,\left( i+1 \right)},\varepsilon_{gdl},P_{agdl,i,\left( i+1 \right)},T_{fc}) \frac{C_{H_{2},agdl,\left( i+1 \right)}-C_{H_{2},agdl,i}}{H_{gdl}/n_{gdl}} $ \label{eq:J_H2_agdl,i,i+1} \\

			$\varepsilon_{gdl} \left[ 1-\texttt{s}_{agdl,n_{gdl}} \right] \frac{d C_{H_{2},agdl,n_{gdl}}}{d t} = \frac{J_{H_{2},agdl,\left( n_{gdl}-1 \right), n_{gdl}} - J_{H_{2},agdl,acl}}{H_{gdl}/n_{gdl}} $ &
			$J_{H_{2},agdl,acl} = - 2D_{a,eff}(\texttt{s}_{agdl,acl},\varepsilon_{agdl,acl},P_{agdl,acl},T_{fc}) \frac{C_{H_{2},acl} - C_{H_2,agdl,n_{gdl}}}{H_{gdl}/n_{gdl}+H_{cl}}   $ \label{eq:J_H2_agdl_acl}  \\ 
			
			$\varepsilon_{cl} \left[ 1-\texttt{s}_{acl} \right] \frac{d C_{H_{2},acl}}{d t} = \frac{J_{H_{2},agdl,acl}}{H_{cl}} + S_{H_{2},acl} $ &
			$S_{H_{2},acl} = - \frac{i_{fc}}{2 F H_{cl}} - \frac{R T_{fc}}{H_{cl} H_{mem}} \left[ k_{H_{2}} \left( \lambda_{mem}, T_{fc} \right) C_{H_{2},acl} + 2 k_{O_{2}} \left( \lambda_{mem}, T_{fc} \right) C_{O_{2},ccl} \right]$ \label{eq:hydrogen_consumption} \\ \hline

			\multicolumn{2}{|c|}{\bf{Oxygen in the GDL and the CL}} \\ \hline
			
			$\forall i \in \llbracket 2, n_{\text{gdl}}-1 \rrbracket :$ &
			$\forall i \in \llbracket 1, n_{\text{gdl}}-1 \rrbracket :$  \\
			
			$\varepsilon_{cl} \left[ 1-\texttt{s}_{ccl} \right] \frac{d C_{O_{2},ccl}}{d t} = \frac{-J_{O_{2},ccl,cgdl}}{H_{cl}} + S_{O_{2},ccl}  $  &
			$S_{O_{2},ccl} = - \frac{i_{fc}}{4 F H_{cl}} - \frac{ R T_{fc}}{H_{cl} H_{mem}} \left[ k_{O_{2}} \left( \lambda_{mem}, T_{fc} \right) C_{O_{2},ccl} + \frac{k_{H_{2}} \left( \lambda_{mem}, T_{fc} \right)}{2} C_{H_{2},acl} \right]$ \label{eq:oxygen_consumption} \\

			$\varepsilon_{gdl} \left[ 1-\texttt{s}_{cgdl,1} \right] \frac{d C_{O_{2},cgdl,1}}{d t} = \frac{J_{O_{2},ccl,cgdl}-J_{O_{2},cgdl,1,2}}{H_{gdl}/n_{gdl}} $ &
			$J_{O_{2},ccl,cgdl} = - 2D_{c,eff}(\texttt{s}_{ccl,cgdl},\varepsilon_{ccl,cgdl},P_{ccl,cgdl},T_{fc}) \frac{C_{O_{2},cgdl,1} - C_{O_{2},ccl}}{H_{gdl}/n_{gdl}+H_{cl}} $ \label{eq:J_O2_ccl_cgdl} \\ 
			
			$\varepsilon_{gdl} \left[ 1-\texttt{s}_{cgdl,i} \right] \frac{d C_{O_{2},cgdl,i}}{d t} = \frac{J_{O_{2},cgdl,\left( i-1 \right),i}-J_{O_{2},cgdl,i,\left( i+1 \right)}}{H_{gdl}/n_{gdl}} $ &
			$J_{O_{2},cgdl,i,\left( i+1 \right)} = - D_{c,eff}(\texttt{s}_{cgdl,i,\left( i+1 \right)},\varepsilon_{gdl},P_{cgdl,i,\left( i+1 \right)},T_{fc}) \frac{C_{O_{2},cgdl,\left( i+1 \right)} - C_{O_{2},cgdl,i}}{H_{gdl}/n_{gdl}}   $ \label{eq:J_O2_cgdl_i_i+1}   \\ 
			
			$\varepsilon_{gdl} \left[ 1-\texttt{s}_{cgdl,n_{gdl}} \right] \frac{d C_{O_{2},cgdl,n_{gdl}}}{d t} = \frac{J_{O_{2},cgdl,\left( n_{gdl}-1 \right),n_{gdl}}-J_{O_{2},cgdl,cgc}}{H_{gdl}/n_{gdl}} $ &
			$J_{O_{2},cgdl,cgc} = h_{c}(P_{cgdl,cgc},T_{fc}) \left[ C_{O_{2},cgdl,n_{gdl}} - C_{O_{2},cgc} \right] $ \label{eq:J_O2_cgdl_cgc}  \\ \hline

		\end{tabularx}
		\caption{Synthesis of the differential equations and the associated matter transport expressions in the stack \cite{gassCriticalReviewProton2024} (2/2)}
		\label{table:synthesis_flows_equadif_2}
	\end{table}
\end{landscape}

\begin{landscape}
	\begin{table}[H]
		\centering
		\small
		\setlength{\abovedisplayskip}{-4pt} 
		\setlength{\belowdisplayskip}{-4pt} 
		\begin{tabularx}{\linewidth}{|YY|} \hline
			
			\multicolumn{2}{|>{\hsize=\dimexpr2\hsize+2\tabcolsep+\arrayrulewidth\relax}Y|}{
				\bf{Coefficients associated to the dissolved water in the membrane}} \\ \hline
			
			\begin{equation}
				a_{w} (C_v,\texttt{s}) = \frac{C_v}{C_{v,sat}} + 2 \texttt{s} \label{eq:water_activity_simplified}
			\end{equation}	
			& \begin{equation}
				D (\lambda) = 4.1 \times 10^{-10} \left[ \frac{\lambda}{25.0} \right]^{0.15} \left[ 1.0 + \tanh\left( \frac{\lambda-2.5}{1.4} \right)\right]
				\label{eq:coef_dif_Kulikovsky}
			\end{equation} \\ 
			
			\multicolumn{2}{|>{\hsize=\dimexpr2\hsize+2\tabcolsep+\arrayrulewidth\relax}Y|}{
				\begin{equation}
					\lambda_{e q}^{cl} = 
					\frac{1}{2} \left( 0.300 + 10.8a_{w} - 16.0a_{w}^{2} + 14.1a_{w}^{3} \right) \cdot \left( 1 - \tanh\left[ 100 \left( a_{w} - 1 \right)\right]\right) + \frac{1}{2} \left( 9.2 +8.6 \left( 1 - \exp\left[ -K_{\text{shape}} \left( a_{w} - 1        \right)\right]\right)\right) \cdot \left( 1 + \tanh \left[ 100 \left( a_{w} - 1 \right)\right]\right) 
			\end{equation}} \\
			
			\begin{equation}
				f_{v}(\lambda) = \frac{ \lambda V_{w} }{ V_{\text{mem}} + \lambda V_{w} } \label{eq:j_sorp_Ge_fv}
			\end{equation}
			& \begin{equation}
				\gamma_{sorp} (\lambda,T_{fc}) = 
				\begin{cases}
					\frac{1.14 \cdot 10^{-5} f_{v}(\lambda)}{H_{cl}} e^{ 2416 \left[ \frac{1}{303} - \frac{1}{T_{f c}} \right] }, & \text{\footnotesize absorption flow} \\
					\frac{4.59 \cdot 10^{-5} f_{v}(\lambda)}{H_{cl}} e^{ 2416 \left[ \frac{1}{303} - \frac{1}{T_{f c}} \right] }, & \text{\footnotesize desorption flow}
				\end{cases}  \label{eq:j_sorp_Ge_ksorp}
			\end{equation}    \\ \hline

			\multicolumn{2}{|>{\hsize=\dimexpr2\hsize+2\tabcolsep+\arrayrulewidth\relax}Y|}{
				\bf{Coefficients associated to liquid water in the GDL and the CL}} \\  \hline
			
			\begin{equation}
				K_{0}(\varepsilon) = \frac{\varepsilon}{8 \ln \left( \varepsilon \right)^{2}} \frac{\left[ \varepsilon - \varepsilon_{p} \right]^{\alpha + 2} r_{f}^{2}} {\left[ 1 - \varepsilon_{p} \right]^{\alpha} \left[ \left[ \alpha + 1 \right]  \varepsilon - \varepsilon_{p} \right]^{2}} e^{ \beta_{1} \varepsilon_{c}} \label{eq:intrinsic_permeability}
			\end{equation} &
			\begin{equation}
				\sigma(T_{fc}) = 235.8 \times 10^{-3} \left[ \frac{647.15 - T_{fc}}{647.15} \right]^{1.256} \left[ 1 - 0.625 \frac{647.15 - T_{fc}}{647.15} \right]
				\label{eq:surface_tension}
			\end{equation} \\ \hline

			\multicolumn{2}{|>{\hsize=\dimexpr2\hsize+2\tabcolsep+\arrayrulewidth\relax}Y|}{
				\bf{Coefficients associated to vapor inside the GDL and the CL}} \\  \hline
			
			\begin{equation}
				\begin{cases}
					& D_{a,eff} \left( \texttt{s},\varepsilon,P,T_{fc} \right)  = 
					\varepsilon \left[ \frac{\varepsilon - \varepsilon_{p}} {1 - \varepsilon_{p}} \right]^{\alpha} \left[ 1 - \texttt{s} \right]^{2}  e^{ \beta_{2} \varepsilon_{c}} D_{a} \left( P,T_{fc} \right) \\
					& D_{c,eff} \left( \texttt{s},\varepsilon,P,T_{fc} \right)  = 
					\varepsilon \left[ \frac{\varepsilon - \varepsilon_{p}} {1 - \varepsilon_{p}} \right]^{\alpha} \left[ 1 - \texttt{s} \right]^{2} e^{ \beta_{2} \varepsilon_{c}} D_{c} \left( P,T_{fc} \right)
				\end{cases}
				\label{eq:effective_diffusion_coefficient}
			\end{equation} &
			\begin{equation}
				\begin{cases}
					& D_{a} \left( P,T_{fc} \right) = 1.644 \cdot 10^{-4} \left[ \frac{T_{fc}}{333} \right]^{2.334} \left[ \frac{101325}{P} \right] \\
					& D_{c} \left( P,T_{fc} \right) = 3.242 \cdot 10^{-5} \left[ \frac{T_{fc}}{333} \right]^{2.334} \left[ \frac{101325}{P} \right]
				\end{cases}
				\label{eq:binary_diffusion_coef_Ohayre} 
			\end{equation} \\	
			
			\begin{equation}
				h_{i}(P,T_{fc}) = S_{h} \frac{D_{i}(P,T_{fc})}{H_{gc}} \text{ $\forall$ i $\in \left\lbrace a,c\right\rbrace$ } \label{eq:Sherwood_number_def}
			\end{equation} & 
			\begin{equation}
				S_{h} = 0.9247 \cdot \ln \left( \frac{W_{gc}}{H_{gc}} \right) + 2.3787 \label{eq:Sherwood_number_expression} 
			\end{equation} \\ \hline
			
			\multicolumn{2}{|>{\hsize=\dimexpr2\hsize+2\tabcolsep+\arrayrulewidth\relax}Y|}{
				\bf{Coefficients associated to $H_{2}$ and $O_{2}$ in the CL}} \\  \hline
			
			\multicolumn{2}{|>{\hsize=\dimexpr2\hsize+2\tabcolsep+\arrayrulewidth\relax}Y|}{
				\begin{equation}
					k_{H_{2}} \left( \lambda, T_{fc} \right)  = 
					\begin{cases}
						\kappa_{co} \left[ 0.29 + 2.2 f_{v}\left( \lambda \right) \right] 10^{-14} \exp \left( \frac{E_{act,H_{2},v}}{R} \left[ \frac{1}{T_{ref}} - \frac{1}{T_{fc}} \right]\right) & if \lambda < 17.6  \\
						\kappa_{co} 1.8 \cdot 10^{-14} \exp \left( \frac{E_{act,H_{2},l}}{R} \left[ \frac{1}{T_{ref}} - \frac{1}{T_{fc}} \right]\right) & if \lambda \geq 17.6 \\
					\end{cases}
					\label{eq:permeability_coefficients_crossover_H2}
			\end{equation}} \\
			
			\multicolumn{2}{|>{\hsize=\dimexpr2\hsize+2\tabcolsep+\arrayrulewidth\relax}Y|}{
				\begin{equation}
					k_{O_{2}} \left( \lambda, T_{fc} \right) = 
					\begin{cases}
						\kappa_{co} \left[ 0.11 + 1.9 f_{v}\left( \lambda \right) \right] 10^{-14} \exp \left( \frac{E_{act,O_{2},v}}{R} \left[ \frac{1}{T_{ref}} - \frac{1}{T_{fc}} \right]\right) & if \lambda < 17.6  \\
						\kappa_{co} 1.2 \cdot 10^{-14} \exp \left( \frac{E_{act,O_{2},l}}{R} \left[ \frac{1}{T_{ref}} - \frac{1}{T_{fc}} \right]\right) & if \lambda \geq 17.6
					\end{cases}
					\label{eq:permeability_coefficients_crossover_O2}
			\end{equation}} \\  \hline
			
		\end{tabularx}
		\caption{Synthesis of the flow coefficients \cite{gassCriticalReviewProton2024}}
		\label{table:synthesis_flow_coef}
	\end{table}
\end{landscape}

\begin{table}[H]
	\centering
	\small
	\setlength{\abovedisplayskip}{-4pt} 
	\setlength{\belowdisplayskip}{-4pt} 
	\begin{tabularx}{13cm}{|YYYYY|} \hline
		
		\textbf{Symbol} &
		\multicolumn{3}{|>{\hsize=\dimexpr3\hsize+3\tabcolsep+\arrayrulewidth\relax}Y|}{
			\textbf{Name (Unit)}} 
		& \textbf{Value} \\ \hline
		
		\multicolumn{5}{|c|}{\bf{Cell model parameters}} \\ \hline
		
		$\rho_{mem}$ &
		\multicolumn{3}{|>{\hsize=\dimexpr3\hsize+3\tabcolsep+\arrayrulewidth\relax}Y|}{
			Density of the dry membrane $(kg.m^{-3})$} 
		& $1980$ \\
		
		$M_{eq}$ &
		\multicolumn{3}{|>{\hsize=\dimexpr3\hsize+3\tabcolsep+\arrayrulewidth\relax}Y|}{
			Equivalent molar mass of ionomer $(kg.mol^{-1})$} 
		& $1.1$ \\
		
		$\varepsilon_{cl}$ &
		\multicolumn{3}{|>{\hsize=\dimexpr3\hsize+3\tabcolsep+\arrayrulewidth\relax}Y|}{
			Porosity of the catalyst layer} 
		& $0.25$ \\
		
		$\theta_{c,gdl}$ &
		\multicolumn{3}{|>{\hsize=\dimexpr3\hsize+3\tabcolsep+\arrayrulewidth\relax}Y|}{
			Contact angle of GDL for liquid water (rad)} 
		& $\frac{2}{3} \pi$ ($120$°) \\
		
		$\theta_{c,cl}$ &
		\multicolumn{3}{|>{\hsize=\dimexpr3\hsize+3\tabcolsep+\arrayrulewidth\relax}Y|}{
			Contact angle of CL for liquid water (rad)} 
		& $1.66$ ($95$°) \\
		
		$\gamma_{cond}$ &
		\multicolumn{3}{|>{\hsize=\dimexpr3\hsize+3\tabcolsep+\arrayrulewidth\relax}Y|}{
			Overall condensation rate constant for water $(s^{-1})$} 
		& $5 \cdot 10^{3}$ \\
		
		$\gamma_{evap}$ &
		\multicolumn{3}{|>{\hsize=\dimexpr3\hsize+3\tabcolsep+\arrayrulewidth\relax}Y|}{
			Overall evaporation rate constant for water $(Pa^{-1}.s^{-1})$} 
		& $10^{-4}$ \\
		
		$K_{shape}$ &
		\multicolumn{3}{|>{\hsize=\dimexpr3\hsize+3\tabcolsep+\arrayrulewidth\relax}Y|}{
			Mathematical factor governing $\lambda_{eq}$ smoothing} 
		& $2$ \\ \hline
		
		\multicolumn{5}{|c|}{\bf{Physical constants}} \\ \hline
		
		$F$ &
		\multicolumn{3}{|>{\hsize=\dimexpr3\hsize+3\tabcolsep+\arrayrulewidth\relax}Y|}{
			Faraday constant $(C.mol^{-1})$} 
		& $96485$ \\ 
		
		$R$ &
		\multicolumn{3}{|>{\hsize=\dimexpr3\hsize+3\tabcolsep+\arrayrulewidth\relax}Y|}{
			Universal gas constant $(J.mol^{-1}.K^{-1})$} 
		& $8.314$ \\ \hline
		
	\end{tabularx}
	\caption{Synthesis of the general parameters for the cell modeling \cite{gassCriticalReviewProton2024}}
	\label{table:synthesis_cell_parameters}
\end{table}
\section{Balance of plant modeling of a PEMFC system}
\label{sec:auxiliary_modeling}

\subsection{An anodic recirculation PEMFC system}

\begin{figure}
	\centering
	\includegraphics[width=16.5cm]{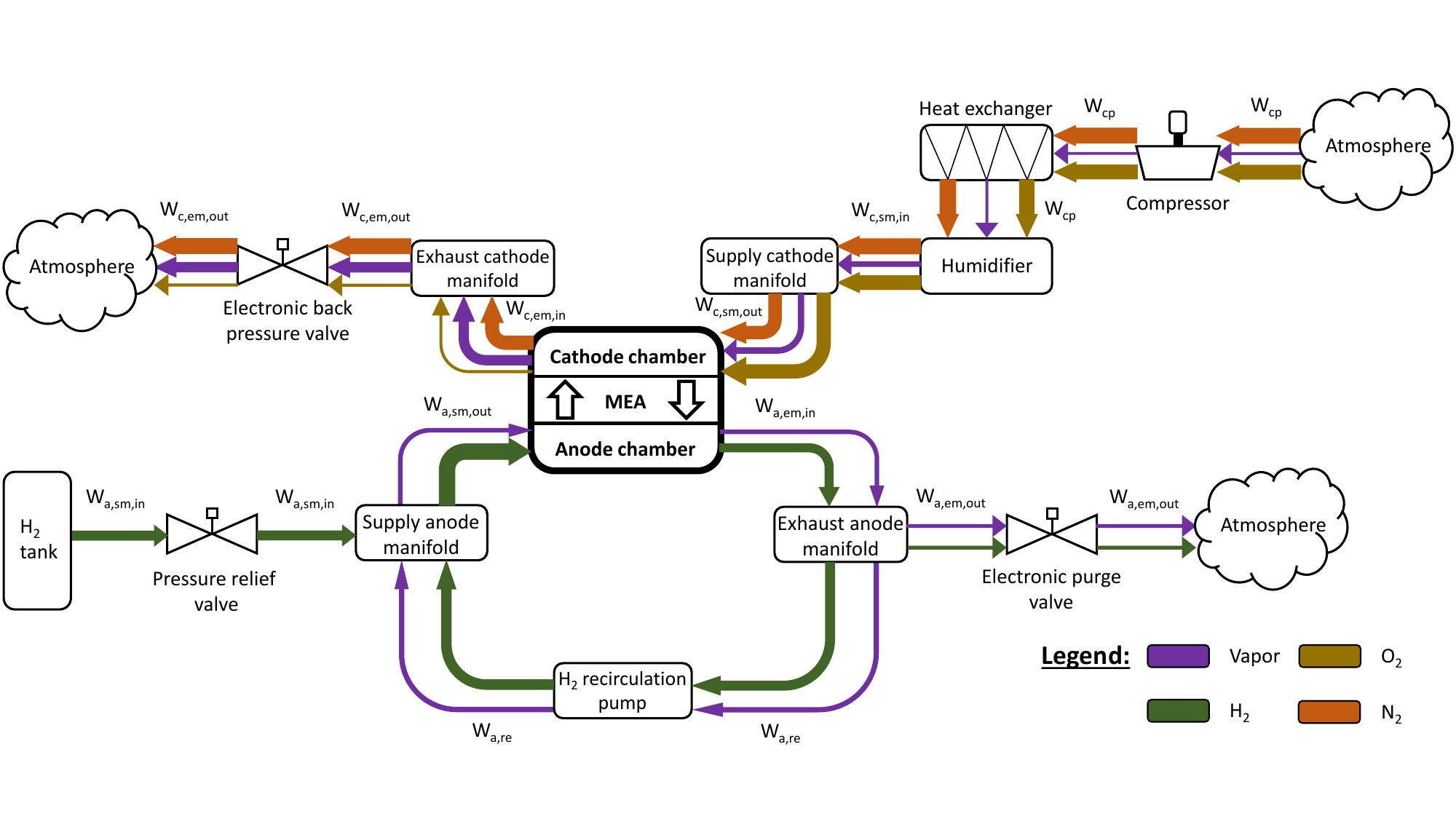}
	\caption{The studied simplified structure of a PEMFC system, consisting of a forced-convective cathode and anodic recirculation.}
	\label{fig:fuel_cell_system}
\end{figure}

In this study, the focus was on considering a fuel cell system rather than examining a single cell only. This approach enables the observation of the auxiliary components' impact on the fuel cells' internal states and performance, which is crucial for control design. Within this investigation, a conventional fuel cell system for vehicles is studied and depicted in figure \ref{fig:fuel_cell_system}.
Specifically, on the anode side, there is a hydrogen storage tank where $H_2$ is maintained at a desired temperature $T$. It is connected to a pressure relief valve that delivers pure $H_2$ to the supply manifold of the anodic chamber. At the outlet of this chamber, there is an exhaust manifold connected both to an electronic purge valve and to a pump that recirculates $H_2$ back to the supply anode manifold. On the cathode side, a compressor supplies ambient air to the stack, passing successively through a heat exchanger, a humidifier, and a supply cathode manifold. At the outlet of the cathodic chamber, an exhaust manifold is directly linked to an electronic back pressure valve. Finally, this valve releases the gases into the atmosphere, without recovering heat or water from the exhaust air.

Thus, with this setup, the fuel cell can be controlled by the user. On the anode side, the inlet pressure is regulated by the pressure relief valve, and the inlet flow by the recirculation pump. It is also assumed here that the hydrogen within its reservoir is maintained at the desired temperature. On the cathode side, the temperature and humidity of the incoming gases are controlled through the heat exchanger and the humidifier. The compressor dictates an inlet flow, and the back-pressure valve regulates the pressure within the cell. 

\textbf{Remark}: This configuration is a simplified version of the one predominantly employed in embedded applications. Yet, during the model validation phase, a modified anode gas supply configuration, similar to the cathode, is utilized to have more flexible control over the operating conditions. This approach is frequently employed in laboratory settings.

\subsection{A 0D, dynamic and isothermal model of the auxiliary system}

For this study, the aim is not to extensively model the auxiliary system. A simple approach is proposed, based on the foundational work of Pukrushpan et al. \cite{pukrushpanControlOrientedModelingAnalysis2004} and in line with the works of Liangfei Xu et al. \cite{xuRobustControlInternal2017}, Y. Shao et al. \cite{shaoComparisonSelfhumidificationEffect2020} and Ling Xu et al. \cite{xuReduceddimensionDynamicModel2021}. These works already provide a clear explanation of the auxiliaries' modeling, and readers are encouraged to refer to them for a more detailed understanding. In the present article, the equations derived from these works are directly adopted, with some additions and modifications detailed below. It is worth noting that the mathematical quantity describing the material flows in auxiliaries is traditionally denoted as W and is in $kg.s^{-1}$, unlike the flows in the cells which are traditionally denoted as J, are calculated per area, and primarily molar ($mol.m^{-2}.s^{-1}$). 

Several simplifying assumptions have been considered here for the modeling of this auxiliary system:
\begin{itemize}[noitemsep] 
	\item Each of the mentioned components is modeled in 0D, meaning the internal parameters in each component are homogeneous.
	\item The current model is isothermal, implying that the temperature $T_{fc}$ is assumed constant throughout the fuel cell system. Thus, the heat exchanger is disregarded here. This assumption is significant, but is expected to be eliminated in future works.
	\item Pressure losses along fuel cell gas channels are not modeled.
	\item The liquid water separator is not modeled. It is assumed that water droplets evacuate so rapidly and efficiently that they do not exist in the auxiliaries. Similarly, any condensation within the auxiliary components is presumed to be promptly removed.
	\item The $H_2$ tank and its pressure relief valve are not directly modeled. It is assumed that this reservoir is infinite, and its valve is perfectly regulated to continuously produce a flow at a constant controlled pressure $P_{a,des}$ at the inlet of the supply anode manifold. 
	\item The electronic purge valve is inactive in this study and so $k_{purge}=0$ in \eqref{eq:Wa_em_out}. 
	\item The dynamic behavior of the compressor and humidifier is simplified at first order considering the desired steady-state flows $W_{cp,des}$ and $W_{c,inj,des}$, along with the time constants $\tau_{cp}$ and $\tau_{hum}$.
	\item It is assumed that the pressure at the compressor outlet equals the pressure in the supply manifold of the cathode: $P_{cp} = P_{csm}$.
	\item It is considered that the recirculation pump reaches its steady state instantly, being much faster than other devices. 
	
\end{itemize}

Certain additions have also been made compared to the existing auxiliary models, such as calculating the humidities in the manifolds (following the same principles stated by Pukrushpan et al. \cite{pukrushpanControlOrientedModelingAnalysis2004}) and controlling the back pressure valve to set the pressure in the stack. The cathode back pressure valve is modeled using a proportional derivative controller as shown in \eqref{eq:pressurevalve}. This is an original idea presented in this paper. The throttle area of this valve, denoted as $A_{bp,c}$, with $A_{bp,c}\in\left[ 0, A_T \right]$, is controlled to affect the quantity of matter exiting the cell and thereby influencing the upstream pressure, $P_{cgc}$ here. 
Then, the proportionality constant $K_p$ is set by considering that the valve takes two seconds to fully open or close and that, during this period, the pressure can change by 0.1 bar. The derivative constant $K_d$ is obtained empirically. 

The linearization of several flows is employed in \eqref{eq:Wa_sm_in}, \eqref{eq:Wa_sm_out}, \eqref{eq:Wa_em_in}, \eqref{eq:Wc_sm_out} and \eqref{eq:Wc_em_in}. 
The exhaust manifolds outflows, in \eqref{eq:Wa_em_out} and \eqref{eq:Wc_em_out}, are not linearized because the pressure difference between the interior of the fuel cell system and the external environment can be significant. Additionally, it has been assumed here that the outflow is necessarily subcritical to avoid the additional instability associated with the piecewise-defined function proposed by Pukrushpan et al. \cite{pukrushpanControlOrientedModelingAnalysis2004}. Furthermore, it should be noted that there are persistent errors in the literature regarding these equations, specifically the omission of the molar mass under the square root and the confusion between sonic and supersonic flows \cite{pukrushpanControlOrientedModelingAnalysis2004, xuRobustControlInternal2017, xuReduceddimensionDynamicModel2021}. Finally, in these equations, $\gamma_{H_2}$ and $\gamma_a$ are considered constants, their value changing only slightly with the alteration of flow composition. 

Knowing these hypotheses and based on the previously mentioned works \cite{xuReduceddimensionDynamicModel2021,pukrushpanControlOrientedModelingAnalysis2004,xuRobustControlInternal2017, shaoComparisonSelfhumidificationEffect2020}, it is possible to construct the system of differential equations, presented in table \ref{table:synthesis_flows_equadif_auxiliary}, which describes the studied auxiliary system. Additionally, the molar masses equations and the balance of plant parameters are provided in tables \ref{table:synthesis_molar_mass_equations} and \ref{table:synthesis_auxiliary_parameters}.

\begin{table} [H]
	\centering
	\small
	\setlength{\abovedisplayskip}{-4pt} 
	\setlength{\belowdisplayskip}{-4pt} 
	\begin{tabularx}{15cm}{|Y|} \hline
		
		\bf{Molar masses equations} \\ \hline
		
		\begin{equation}
			M_{asm} = \frac{\Phi_{asm} P_{sat}\left(T_{fc}\right)}{P_{asm}} M_{H_2O} + \frac{P_{asm} - \Phi_{asm} P_{sat}\left(T_{fc}\right)}{P_{asm}} M_{H_2}
		\end{equation} \\
		
		\begin{equation}
			M_{aem} = \frac{\Phi_{aem} P_{sat}\left(T_{fc}\right)}{P_{aem}} M_{H_2O} + \frac{P_{aem} - \Phi_{aem} P_{sat}\left(T_{fc}\right)}{P_{aem}} M_{H_2}
		\end{equation} \\ 
		
		\begin{equation}
			M_{csm} = \frac{\Phi_{csm} P_{sat}\left(T_{fc}\right)}{P_{csm}} M_{H_2O} + y_{O_2,ext} \frac{P_{csm} - \Phi_{csm} P_{sat}\left(T_{fc}\right)}{P_{csm}} M_{O_2} + \left( 1 - y_{O_2,ext} \right) \frac{P_{csm} - \Phi_{csm} P_{sat}\left(T_{fc}\right)}{P_{csm}} M_{N_2}
		\end{equation} \\ 
		
		\begin{equation}
			M_{agc} = \frac{\Phi_{agc} P_{sat}\left(T_{fc}\right)}{P_{agc}} M_{H_2O} + \frac{P_{agc} - \Phi_{agc} P_{sat}\left(T_{fc}\right)}{P_{agc}} M_{H_2}
		\end{equation} \\ 
		
		\begin{equation}
			M_{i} = \frac{\Phi_{i} P_{sat}\left(T_{fc}\right)}{P_{i}} M_{H_2O} + y_{O_2,i} \frac{P_{i} - \Phi_{i} P_{sat}\left(T_{fc}\right)}{P_{i}} M_{O_2} + \left( 1 - y_{O_2,i} \right) \frac{P_{i} - \Phi_{cgc} P_{sat}\left(T_{fc}\right)}{P_{i}} M_{N_2}, i \in \left\lbrace cem, cgc, ext \right\rbrace 
		\end{equation} \\ \hline
		
	\end{tabularx}
	\caption{Synthesis of the molar masses equations}
	\label{table:synthesis_molar_mass_equations}
\end{table}

\begin{table}[H]
	\centering
	\small
	\setlength{\abovedisplayskip}{-4pt} 
	\setlength{\belowdisplayskip}{-4pt} 
	\begin{tabularx}{17cm}{|YYYYY|} \hline
		
		\textbf{Symbol} &
		\multicolumn{3}{|>{\hsize=\dimexpr3\hsize+3\tabcolsep+\arrayrulewidth\relax}Y|}{
			\textbf{Name (Unit)}} 
		& \textbf{Value} \\ \hline
		
		\multicolumn{5}{|c|}{\bf{Auxiliary system model parameters}} \\ \hline
		
		$\tau_{cp}$ &
		\multicolumn{3}{|>{\hsize=\dimexpr3\hsize+3\tabcolsep+\arrayrulewidth\relax}Y|}{
			Air compressor time constant $(s)$} 
		& $1$ \cite{xuRobustControlInternal2017} \\
		
		$\tau_{hum}$ &
		\multicolumn{3}{|>{\hsize=\dimexpr3\hsize+3\tabcolsep+\arrayrulewidth\relax}Y|}{
			Humidifier time constant $(s)$} 
		& $5$ \cite{xuRobustControlInternal2017} \\
		
		$K_{p}$ &
		\multicolumn{3}{|>{\hsize=\dimexpr3\hsize+3\tabcolsep+\arrayrulewidth\relax}Y|}{
			Proportionality constant of the back pressure valve controller $(m^{2}.s^{-1}.Pa^{-1})$} 
		& $5 \cdot 10^{-8}$ \\
		
		$K_{d}$ &
		\multicolumn{3}{|>{\hsize=\dimexpr3\hsize+3\tabcolsep+\arrayrulewidth\relax}Y|}{
			Derivative constant of the back pressure valve controller $(m^{2}.Pa^{-1})$} 
		& $10^{-8}$ \\
		
		$C_D$ &
		\multicolumn{3}{|>{\hsize=\dimexpr3\hsize+3\tabcolsep+\arrayrulewidth\relax}Y|}{
			Throttle discharge coefficient} 
		& $0.05$ \cite{xuRobustControlInternal2017} \\
		
		$k_{sm,in}$ &
		\multicolumn{3}{|>{\hsize=\dimexpr3\hsize+3\tabcolsep+\arrayrulewidth\relax}Y|}{
			Nozzle orifice coefficient at the inlet supply manifold $(kg.Pa^{-1}.s^{-1})$} 
		& $1.0 \cdot 10^{-5}$ \\
		
		$k_{sm,out}$ &
		\multicolumn{3}{|>{\hsize=\dimexpr3\hsize+3\tabcolsep+\arrayrulewidth\relax}Y|}{
			Nozzle orifice coefficient at the outlet supply manifold $(kg.Pa^{-1}.s^{-1})$} 
		& $8.0 \cdot 10^{-6}$ \cite{xuRobustControlInternal2017} \\ \hline
		
		\multicolumn{5}{|c|}{\bf{Auxiliary system physical parameters}} \\ \hline
		
		$n_{cell}$ &
		\multicolumn{3}{|>{\hsize=\dimexpr3\hsize+3\tabcolsep+\arrayrulewidth\relax}Y|}{
			Number of cells inside the stack} 
		& $5$ \\
		
		$V_{sm}$ &
		\multicolumn{3}{|>{\hsize=\dimexpr3\hsize+3\tabcolsep+\arrayrulewidth\relax}Y|}{
			Supply manifold volume $(m^{3})$} 
		& $7.0 \cdot 10^{-3}$ \cite{xuRobustControlInternal2017} \\
		
		$V_{em}$ &
		\multicolumn{3}{|>{\hsize=\dimexpr3\hsize+3\tabcolsep+\arrayrulewidth\relax}Y|}{
			Exhaust manifold volume $(m^{3})$} 
		& $2.4 \cdot 10^{-3}$ \cite{xuRobustControlInternal2017} \\ 
		
		$A_T$ &
		\multicolumn{3}{|>{\hsize=\dimexpr3\hsize+3\tabcolsep+\arrayrulewidth\relax}Y|}{
			Exhaust manifold throttle area $(m^{2})$} 
		& $1.18 \cdot 10^{-3}$ \cite{xuRobustControlInternal2017} \\ \hline
		
		\multicolumn{5}{|c|}{\bf{Physical constants}} \\ \hline
		
		$\gamma_{H_2}$ &
		\multicolumn{3}{|>{\hsize=\dimexpr3\hsize+3\tabcolsep+\arrayrulewidth\relax}Y|}{
			Heat capacity ratio of $H_2$ at 100°C} 
		& $1.404$ \\
		
		$\gamma_a$ & 
		\multicolumn{3}{|>{\hsize=\dimexpr3\hsize+3\tabcolsep+\arrayrulewidth\relax}Y|}{
			Heat capacity ratio of dry air at 100°C} 
		& $1.401$ \\
		
		$M_{H_2}$ & 
		\multicolumn{3}{|>{\hsize=\dimexpr3\hsize+3\tabcolsep+\arrayrulewidth\relax}Y|}{
			Molar mass of $H_2$ $(kg.mol^{-1})$} 
		& $2 \cdot 10^{-3}$ \\
		
		$M_{H_2O}$ &
		\multicolumn{3}{|>{\hsize=\dimexpr3\hsize+3\tabcolsep+\arrayrulewidth\relax}Y|}{
			Molar mass of $H_2O$ $(kg.mol^{-1})$} 
		& $1.8 \cdot 10^{-2}$ \\
		
		$M_{O_2}$ & 
		\multicolumn{3}{|>{\hsize=\dimexpr3\hsize+3\tabcolsep+\arrayrulewidth\relax}Y|}{
			Molar mass of $O_2$ $(kg.mol^{-1})$} 
		& $3.2 \cdot 10^{-2}$ \\
		
		$M_{N_2}$ & 
		\multicolumn{3}{|>{\hsize=\dimexpr3\hsize+3\tabcolsep+\arrayrulewidth\relax}Y|}{
			Molar mass of $N_2$ $(kg.mol^{-1})$} 
		& $2.8 \cdot 10^{-2}$ \\ \hline
		
		\multicolumn{5}{|c|}{\bf{External environmental parameters}} \\ \hline
		
		$T_{ext}$ &
		\multicolumn{3}{|>{\hsize=\dimexpr3\hsize+3\tabcolsep+\arrayrulewidth\relax}Y|}{
			Outside temperature $(K)$} 
		& $298$ \\ 
		
		$P_{ext}$ &
		\multicolumn{3}{|>{\hsize=\dimexpr3\hsize+3\tabcolsep+\arrayrulewidth\relax}Y|}{
			Outside pressure $(Pa)$} 
		& $101325$ \\ 
		
		$\Phi_{ext}$ &
		\multicolumn{3}{|>{\hsize=\dimexpr3\hsize+3\tabcolsep+\arrayrulewidth\relax}Y|}{
			Outside relative humidity} 
		& $0.4$ \\ 
		
		$y_{O_2,ext}$ &
		\multicolumn{3}{|>{\hsize=\dimexpr3\hsize+3\tabcolsep+\arrayrulewidth\relax}Y|}{
			Molar fraction of $O_2$ in ambient dry air} 
		& $0.2095$ \\ \hline
		
	\end{tabularx}
	\caption{Synthesis of the necessary parameters for the balance of plant modeling}
	\label{table:synthesis_auxiliary_parameters}
\end{table}

\begin{landscape}
	\begin{table}
		\centering
		\small
		\setlength{\abovedisplayskip}{-7pt} 
		\setlength{\belowdisplayskip}{-5pt} 
		\begin{tabularx}{\linewidth}{|Y|Y|} \hline		
			
			\bf{Dynamical models}
			& \bf{Matter flow expressions} \\ \hline \hline
			
			\multicolumn{2}{|c|}{\bf{Manifolds at the anode}} \\ \hline
			
			\multirow{3.5}{*}{
				\begin{minipage}{0.98\linewidth}
					\begin{equation}
						\frac{d P_{asm}}{d t} = \frac{R T_{fc}}{V_{sm} M_{asm}} \left[ W_{asm,in} + W_{are} - n_{cell} W_{asm,out} \right] 
					\end{equation}
			\end{minipage}}
			& \begin{equation}
				W_{asm,in} = k_{sm,in} \left[ P_{a,des} - P_{asm} \right] 
				\label{eq:Wa_sm_in}
			\end{equation} \\ 
			
			& \begin{equation}
				W_{v,asm,in} = \frac{\Phi_{aem} P_{sat}\left(T_{fc}\right)}{M_{aem} P_{aem}} W_{are}
				\label{eq:Wv_a_sm_in}
			\end{equation} \\ 
			
			\multirow{3.5}{*}{
				\begin{minipage}{0.98\linewidth}
					\begin{equation}
						\frac{d P_{aem}}{d t} = \frac{R T_{fc}}{V_{em} M_{aem}} \left[ n_{cell} W_{aem,in} - W_{are} - W_{aem,out} \right]
					\end{equation}
			\end{minipage}}
			& \begin{equation}
				W_{asm,out} = k_{sm,out} \left[ P_{asm} - P_{agc} \right] 
				\label{eq:Wa_sm_out}
			\end{equation} \\ 
			
			& \begin{equation}
				W_{are} = n_{cell} M_{aem} \frac{P_{aem}}{P_{aem} - \Phi_{aem} P_{sat}\left( T_{fc} \right)}  \frac{\left[ S
					_{a} - 1 \right] \left[ i_{fc} + i_{n} \right] A_{act}}{2 F} 
			\end{equation} \\ 
			
			\multirow{3.25}{*}{
				\begin{minipage}{0.98\linewidth}
					\begin{equation}
						\frac{d \Phi_{asm}}{d t} = \frac{R T_{fc}}{V_{sm} P_{sat}\left(T_{fc}\right)} \left[ W_{v,asm,in} - J_{v,a,in} H_{gc} W_{gc} n_{cell} \right]
					\end{equation}
			\end{minipage}}
			& \begin{equation}
				W_{aem,in} = k_{em,in} \left[ P_{agc} - P_{aem} \right] 
				\label{eq:Wa_em_in}
			\end{equation} \\ 
			
			\multirow{8}{*}{
				\begin{minipage}{0.98\linewidth}
					\begin{equation}
						\frac{d \Phi_{aem}}{d t} = \frac{R T_{fc}}{V_{em} P_{sat}\left(T_{fc}\right)} \left[ J_{v,a,out} H_{gc} W_{gc} n_{cell} - W_{v,asm,in} - W_{v,aem,out} \right]
					\end{equation}
			\end{minipage}}
			& \begin{equation}
				W_{aem,out} = k_{purge} \frac{C_{D} A_{T} P_{aem}}{\sqrt{R T_{fc}}} \left( \frac{P_{ext}}{P_{aem}} \right)^{\frac{1}{\gamma_{H_{2}}}} \sqrt{M_{agc} \frac{2 \gamma_{H_{2}}}{\gamma_{H_{2}} - 1} \left[ 1 - \left( \frac{P_{ext}}{P_{aem}} \right)^{\frac{\gamma_{H_{2}}-1}{\gamma_{H_{2}}}} \right]} 
				\label{eq:Wa_em_out}
			\end{equation} \\
			
			& \begin{equation}
				W_{v,aem,out} = \frac{\Phi_{aem} P_{sat}\left(T_{fc}\right)}{M_{aem} P_{aem}} W_{aem,out}
				\label{eq:Wv_a_em_out}
			\end{equation} \\ \hline

			\multicolumn{2}{|c|}{\bf{Manifolds at the cathode}} \\ \hline
			
			\multirow{3.5}{*}{
				\begin{minipage}{0.98\linewidth}
					\begin{equation}
						\frac{d P_{csm}}{d t} = \frac{R T_{fc}}{V_{sm} M_{csm}} \left[ W_{csm,in} - n_{cell} W_{csm,out} \right]
					\end{equation}
			\end{minipage}}
			& \begin{equation}
				W_{csm,in} = W_{cp} + W_{c,inj} 
			\end{equation} \\ 
			
			\multirow{6}{*}{
				\begin{minipage}{0.98\linewidth}
					\begin{equation}
						\frac{d P_{cem}}{d t} = \frac{R T_{fc}}{V_{em} M_{cem}} \left[ n_{cell} W_{cem,in} - W_{cem,out} \right] 
					\end{equation}
			\end{minipage}}
			& \begin{equation}
				W_{v,csm,in} = \frac{\Phi_{ext} P_{sat}\left(T_{ext}\right)}{M_{ext} P_{ext}} W_{cp} + \frac{1}{M_{H_{2}O}} W_{c,inj}
				\label{eq:Wv_c_sm_in}
			\end{equation} \\ 
			
			& \begin{equation}
				W_{csm,out} = k_{sm,out} \left[ P_{csm} - P_{cgc} \right] 
				\label{eq:Wc_sm_out}
			\end{equation} \\ 
			
			\multirow{4.5}{*}{
				\begin{minipage}{0.98\linewidth}
					\begin{equation}
						\frac{d \Phi_{csm}}{d t} = \frac{R T_{fc}}{V_{sm} P_{sat}\left(T_{fc}\right)} \left[ W_{v,csm,in} - J_{v,c,in} H_{gc} W_{gc} n_{cell} \right]
					\end{equation}
			\end{minipage}}
			& \begin{equation}
				W_{cem,in} = k_{em,in} \left[ P_{cgc} - P_{cem} \right] 
				\label{eq:Wc_em_in}
			\end{equation} \\ 
			
			\multirow{7.5}{*}{
				\begin{minipage}{0.98\linewidth}
					\begin{equation}
						\frac{d \Phi_{cem}}{d t} = \frac{R T_{fc}}{V_{em} P_{sat}\left(T_{fc}\right)} \left[ J_{v,c,out} H_{gc} W_{gc} n_{cell} - W_{v,cem,out} \right]
					\end{equation}
			\end{minipage}}
			& \begin{equation}
				W_{cem,out} = \frac{C_{D} A_{bp,c} P_{cem}}{\sqrt{R T_{fc}}} \left( \frac{P_{ext}}{P_{cem}} \right)^{\frac{1}{\gamma_a}} \sqrt{M_{cgc} \frac{2 \gamma_a}{\gamma_a - 1} \left[ 1 - \left( \frac{P_{ext}}{P_{cem}} \right)^{\frac{\gamma_a - 1}{\gamma_a}} \right]} 
				\label{eq:Wc_em_out}
			\end{equation} \\
			
			& \begin{equation}
				W_{v,cem,out} = \frac{\Phi_{cem} P_{sat}\left(T_{fc}\right)}{M_{cem} P_{cem}} W_{cem,out}
				\label{eq:Wv_c_em_out}
			\end{equation} \\  \hline

			\multicolumn{2}{|c|}{\bf{Air compressor, humidifiers and back-pressure valve}} \\ \hline
			
			\multirow{3}{*}{
				\begin{minipage}{0.98\linewidth}
					\begin{equation}
						\frac{d W_{cp}}{d t} = \frac{W_{cp,des} - W_{cp}}{\tau_{cp}}
					\end{equation}
			\end{minipage}}
			& \begin{equation}  
				W_{cp,des} = n_{cell} M_{ext} \frac{P_{ext}}{P_{ext} - \Phi_{ext} P_{sat}\left(T_{ext}\right)} \frac{1}{y_{O_{2},ext}} \frac{S_{c} \left[ i_{fc} + i_{n} \right] A_{act}}{4 F} 
			\end{equation} \\
			
			\multirow{3}{*}{
				\begin{minipage}{0.98\linewidth}
					\begin{equation}
						\frac{d W_{c,inj}}{d t} = \frac{W_{c,inj,des} - W_{c,inj}}{\tau_{hum}} 
					\end{equation}
			\end{minipage}}
			& \begin{equation}
				W_{c,inj,des} = W_{c,v,des} - W_{v,hum,in}
			\end{equation} \\
			
			\multirow{5.5}{*}{
				\begin{minipage}{0.98\linewidth}
					\begin{equation}
						\frac{d A_{bp,c}}{d t} = 
						\begin{cases}
							0, & \text{if $A_{bp,c} \geq A_T$ and $\frac{d A_{bp,c}}{d t} > 0$} \\
							0, & \text{if $A_{bp,c} \leq 0$ and $\frac{d A_{bp,c}}{d t} < 0$} \\
							- K_{p} \left[ P_{c,des} - P_{cgc} \right] + K_{d} \frac{d P_{cgc}}{d t}, & \text{else}
						\end{cases}
						\label{eq:pressurevalve}
					\end{equation}
			\end{minipage}}
			& \begin{equation}
				W_{c,v,des} = M_{H_{2}O} \frac{\Phi_{c,des} P_{sat}\left(T_{fc}\right)}{P_{cp}} \frac{W_{cp}}{M_{ext}} 
			\end{equation} \\
			
			& \begin{equation}
				W_{v,hum,in} = M_{H_{2}O} \frac{\Phi_{ext} P_{sat}\left(T_{ext}\right)}{P_{ext}} \frac{W_{cp}}{M_{ext}}
			\end{equation} \\ \hline
			
		\end{tabularx}
		\caption{Synthesis of the differential equations and the associated matter transport expressions in the auxiliary system}
		\label{table:synthesis_flows_equadif_auxiliary}
	\end{table}
\end{landscape}

\subsection{Flaws of this balance of plant model}
\label{subsec:flaws_bop_model}
The model proposed for the auxiliaries has several flaws. First, the only equations in the literature that are practically applicable for calculating the manifold inflow or outflow rates based on a pressure difference are those of the form given in \eqref{eq:Wa_sm_in}, \eqref{eq:Wa_sm_out}, \eqref{eq:Wa_em_in}, \eqref{eq:Wc_sm_out} and \eqref{eq:Wc_em_in}. There are other equations derived from the Bernoulli's principle, such as the one proposed by Pukrushpan \cite{pukrushpanControlOrientedModelingAnalysis2004} and used in \eqref{eq:Wa_em_out} and \eqref{eq:Wc_em_out}. However, these equations, in addition to assuming steady and incompressible flow, which is not valid in the present case, introduce a square root of the pressure difference. This square root function imposes a direction to the flow, as the pressure difference has to be positive, preventing symmetric considerations. This is problematic because, around initial conditions, the flows can be temporarily and briefly reversed. Gas could enter the GC through the outlet manifold, or gas could exit the GC towards the inlet manifold. Square root is also a source of numerical instability when solving the equations. Equations \eqref{eq:Wa_sm_in}, \eqref{eq:Wa_sm_out}, \eqref{eq:Wa_em_in}, \eqref{eq:Wc_sm_out} and \eqref{eq:Wc_em_in}, on the other hand, are obtained by linearizing the aforementioned Bernoulli principle. While it solves the asymmetry issue, the linearization requires that the pressure difference on both sides of the orifice must be very small, which may not be the case in practice. To the best of the authors' knowledge, no superior models for these flows currently exist.

\section{Voltage modeling of a PEM cell}
\subsection{General expressions}

The voltage polarization expressions, based on the authors' previous work \cite{gassCriticalReviewProton2024}, are adapted for this model and given in table \ref{table:synthesis_spotlighted_voltage_expressions}. Two significant scientific additions are noteworthy here: $\kappa_{co}$ and $s_{lim}$. They have been implemented to enable the model to more accurately simulate reality when comparing results with the experimental data. A discussion dedicated to them can be found in sections \ref{subsec:crossover_correction_coefficient} and \ref{subsec:s_lim}. Finally, general parameters for modeling the cell voltage are furnished in table \ref{table:synthesis_cell_voltage_parameters}, while the cell's voltage specific parameters contingent upon the cell type used are delineated in section \ref{sec:validation_model}.

\begin{table}[H]
	\centering
	\small
	\setlength{\abovedisplayskip}{-4pt} 
	\setlength{\belowdisplayskip}{-4pt} 
	\begin{tabularx}{\linewidth}{|Y|YYY|} \hline
		
		\multicolumn{4}{|>{\hsize=\dimexpr4\hsize+8\tabcolsep+3\arrayrulewidth\relax}Y|}{
			\bf{Voltage polarization expressions}} \\ \hline \hline
		
		\multirow{2}{*}{
			\textbf{The apparent voltage}} &
		\multicolumn{3}{>{\hsize=\dimexpr3\hsize+4\tabcolsep+2\arrayrulewidth\relax}Y|}{
			\begin{equation}
				U_{cell} = U_{eq} - \eta_{c} - i_{fc}  \left[ R_{p} + R_{e} \right]
				\label{eq:apparent_voltage}
		\end{equation}} \\ \hline
		
		\multirow{3}{*}{
			\textbf{The equilibrium potential}} &
		\multicolumn{3}{>{\hsize=\dimexpr3\hsize+4\tabcolsep+2\arrayrulewidth\relax}Y|}{
			\begin{equation}
				U_{eq} = E^{0} - 8.5 \cdot 10^{-4} \left[ T_{fc} - 298.15 \right] + \frac{R T_{fc}}{2 F} \left[ \ln \left( \frac{R T_{fc} C_{H_{2},acl}}{P_{\text{ref}}} \right) + \frac{1}{2} \ln \left( \frac{R T_{fc} C_{O_{2},ccl}}{P_{\text{ref}}} \right)\right] \label{eq:equilibrium_voltage}
		\end{equation}} \\ \hline
		
		\multirow{9}{*}{\bf{The overpotential}}
		& \multicolumn{2}{>{\hsize=\dimexpr2\hsize+2\tabcolsep+\arrayrulewidth\relax}Y}{
			\begin{equation}
				\eta_{c} = \frac{1}{f_{drop}\left( \texttt{s},P \right)} \frac{R T_{fc}}{\alpha_{c} F} ln \left( \frac{i_{fc}+i_{n}}{i_{0,c}^{ref}} \left[ \frac{C_{O_{2}}^{ref}}{C_{O_{2},ccl}} \right]^{\kappa_{c}} \right) \label{eq:overpotential}
		\end{equation}}
		& \begin{equation}
			\texttt{s}_{lim}  = a_{\texttt{s}_{lim}} P_{des} + b_{\texttt{s}_{lim}} \label{eq:limit_liquid_saturation}
		\end{equation} \\
		
		& \multicolumn{2}{>{\hsize=\dimexpr2\hsize+2\tabcolsep+\arrayrulewidth\relax}Y}{
			\begin{equation}
				f_{drop} \left( \texttt{s},P \right) = \frac{1}{2} \left[ 1.0 - tanh\left[ \frac{4 \texttt{s}_{ccl} - 2 \texttt{s}_{lim} - 2 \texttt{s}_{switch}}{\texttt{s}_{lim} - \texttt{s}_{switch}} \right]\right]  \label{eq:f_drop}
		\end{equation}}
		& \begin{equation}
			\texttt{s}_{switch}  = a_{switch} \cdot \texttt{s}_{lim} 
			\label{eq:s_switch}
		\end{equation} \\
		
		& \multicolumn{2}{>{\hsize=\dimexpr2\hsize+2\tabcolsep+\arrayrulewidth\relax}Y}{
			\begin{equation}
				\begin{cases}
					i_{co,H_{2}} = 2 F k_{H_{2}} \left( \lambda_{mem}, T_{fc} \right) R T_{fc} C_{H_{2},acl} \\
					i_{co,O_{2}} = 4 F k_{O_{2}} \left( \lambda_{mem}, T_{fc} \right) R T_{fc} C_{O_{2},ccl}
				\end{cases}
				\label{eq:internal_crossover_current_density}
		\end{equation}}
		& 
		\begin{equation}
			i_{n} = i_{co,H_{2}} + i_{co,O_{2}} \label{eq:internal_current_density}
		\end{equation} \\ \hline
		
		\multirow{11}{*}{
			\textbf{The proton resistance}} &
		\multicolumn{3}{>{\hsize=\dimexpr3\hsize+4\tabcolsep+2\arrayrulewidth\relax}Y|}{
			\normalsize
			\begin{equation}
				\setlength{\abovedisplayskip}{-7pt} 
				\setlength{\belowdisplayskip}{-15pt} 
				R_{p} = R_{mem} + R_{ccl}
				\label{eq:R_p}
				\end{equation}} \\	
			
		& \multicolumn{3}{>{\hsize=\dimexpr3\hsize+4\tabcolsep+2\arrayrulewidth\relax}Y|}{
			\normalsize
			\begin{equation}
				\setlength{\abovedisplayskip}{0pt} 
				\setlength{\belowdisplayskip}{0pt} 
				R_{mem} =
				\begin{cases}
					\frac{H_{mem}}{\left[0.5139 \cdot \lambda_{mem} - 0.326 \right] \exp \left( 1268 \left[ \frac{1}{303.15} - \frac{1}{T_{fc}} \right]\right)}, & \text{if } \lambda_{mem} \geq 1 \\
					\frac{H_{mem}}{0.1879 \exp \left( 1268 \left[ \frac{1}{303.15} - \frac{1}{T_{fc}} \right]\right)}, & \text{if } \lambda_{mem} < 1
				\end{cases}
				\label{eq:R_mem}
			\end{equation}} \\
		
		& \multicolumn{3}{>{\hsize=\dimexpr3\hsize+4\tabcolsep+2\arrayrulewidth\relax}Y|}{
			\normalsize
			\begin{equation}
				\setlength{\abovedisplayskip}{0pt} 
				\setlength{\belowdisplayskip}{0pt} 
				R_{ccl} =
				\begin{cases}
					\frac{H_{cl}}{\varepsilon_{mc}^{\tau} \left[0.5139 \cdot \lambda_{ccl} - 0.326 \right] \exp \left( 1268 \left[ \frac{1}{303.15} - \frac{1}{T_{fc}} \right]\right)}, & \text{if } \lambda_{ccl} \geq 1 \\
					\frac{H_{cl}}{0.1879 \varepsilon_{mc}^{\tau} \exp \left( 1268 \left[ \frac{1}{303.15} - \frac{1}{T_{fc}} \right]\right)}, & \text{if } \lambda_{ccl} < 1
				\end{cases} 
		\end{equation}} \\ \hline
		
	\end{tabularx}
	\caption{Synthesis of the voltage polarization expressions}
	\label{table:synthesis_spotlighted_voltage_expressions}
\end{table}

\begin{table}[H]
	\centering
	\small
	\setlength{\abovedisplayskip}{-4pt} 
	\setlength{\belowdisplayskip}{-4pt} 
	\begin{tabularx}{10cm}{|YYYYY|} \hline
		
		\multicolumn{5}{|c|}{\bf{Cell voltage model parameters}} \\ \hline
		
		\textbf{Symbol} &
		\multicolumn{3}{|>{\hsize=\dimexpr3\hsize+3\tabcolsep+\arrayrulewidth\relax}Y|}{
			\textbf{Name (Unit)}} 
		& \textbf{Value} \\ \hline
		
		$C_{O_2,ref}$ &
		\multicolumn{3}{|>{\hsize=\dimexpr3\hsize+3\tabcolsep+\arrayrulewidth\relax}Y|}{
			Reference concentration of $O_2$ $(mol.m^{-3})$} 
		& $3.39$ \\
		
		$\alpha_{c}$ &
		\multicolumn{3}{|>{\hsize=\dimexpr3\hsize+3\tabcolsep+\arrayrulewidth\relax}Y|}{
			Cathode transfer coefficient} 
		& $0.5$ \\
		
		$E_{0}$ &
		\multicolumn{3}{|>{\hsize=\dimexpr3\hsize+3\tabcolsep+\arrayrulewidth\relax}Y|}{
			Standard-state reversible voltage $(V)$} 
		& $1.229$ \\
		
		$P_{ref}$ &
		\multicolumn{3}{|>{\hsize=\dimexpr3\hsize+3\tabcolsep+\arrayrulewidth\relax}Y|}{
			Reference pressure $(Pa)$} 
		& $10^{5}$ \\
		
		$E_{act}$ &
		\multicolumn{3}{|>{\hsize=\dimexpr3\hsize+3\tabcolsep+\arrayrulewidth\relax}Y|}{
			Activation energy $(J.mol^{-1})$} 
		& $73.2 \cdot 10^{3}$ \\ \hline

	\end{tabularx}
	\caption{Synthesis of the general parameters for the cell voltage modeling \cite{gassCriticalReviewProton2024}}
	\label{table:synthesis_cell_voltage_parameters}
\end{table}

\subsection{New parameter: the crossover correction coefficient $\kappa_{co}$}
\label{subsec:crossover_correction_coefficient}

Expressing the crossover of reactants in fuel cell models is useful for several reasons. First, it is essential for accurately considering the open-circuit voltage in cells and thus obtaining a proper representation of the polarization curve. Furthermore, this information could be valuable to the operator in cases where the cell is temporarily idle. In fact, it is possible to assess the need to flush the anode of any remaining hydrogen, which can lead to cell degradation, when the shutdown is brief and so the quantity of material crossing the membrane is potentially not significant. In such cases, a decision must be made between degradation resulting from purging with ambient air and degradation arising from material crossover.

However, the most notable mathematical expression in the literature, which characterizes this phenomenon, dates back to 2004 \cite{weberTransportPolymerElectrolyteMembranes2004}, as discussed in the authors' previous work \cite{gassCriticalReviewProton2024}. According to the authors' results, this expression is not sufficient to describe the complexity of the crossover in recent stacks. To address this issue, and while awaiting further experiments, the authors propose adding a corrective parameter, denoted here as $\kappa_{co}$, to the permeability coefficients of hydrogen and oxygen in the membrane $\kappa_{H_{2}}$ and $\kappa_{O_{2}}$. This modification has been directly incorporated into equations \eqref{eq:permeability_coefficients_crossover_H2} and \eqref{eq:permeability_coefficients_crossover_O2}. $\kappa_{co}$ is undetermined and requires calibration to be identified for the specific stack under investigation. Further details on the calibration stage are discussed in section \ref{sec:validation_model}.

\subsection{New physical quantity: the limit liquid water saturation coefficient $s_{lim}$}
\label{subsec:s_lim}

To the authors' knowledge, current models struggle to physically incorporate concentration drop during the simulation of polarization curves. Thus, the most commonly used approach so far does not leverage the fuel cell's physics to explain this drop, but rather involves artificially introducing a new element into the equations. For instance,  \eqref{eq:U_conc} is a widely known equation which has been used to quantify the concentration voltage loss \cite{dicksFuelCellSystems2018, ohayreFuelCellFundamentals2016,pukrushpanControlOrientedModelingAnalysis2004,yangEffectsOperatingConditions2019,santarelliParametersEstimationPEM2006,williamsAnalysisPolarizationCurves2005}. In this equation, $i_{lim}$ is introduced to define the limit current density at which the concentration drop becomes inevitable. In most studies, $i_{lim}$ is commonly considered as a constant. However, operational conditions invariably influence its value, consequently altering the current density level at which the concentration drop manifests. $i_{lim}$ therefore should be regarded as a function of the operational conditions, for which the link has yet to be identified. 

\begin{equation}
	U_{conc} = \frac{RT}{2F} ln\left( \frac{i_{lim}}{i_{lim} - i_{fc}} \right)
	\label{eq:U_conc}
\end{equation}

Next, it is necessary to clarify the use of the coefficient $i_{lim}$ for modeling purposes. As soon as more complex models than lumped-parameter models are employed and internal state data within the catalytic layers are available, the physical representation of $i_{lim}$ changes. It ceases to remain the sole mathematical element in the voltage equations that delineate concentration losses arising from gas diffusion limitations within the cell. This limitation occurs when the concentration of oxygen or hydrogen drops to zero within their respective catalytic layers, and the physical and operating conditions do not permit further supply to this region to counterbalance material consumption at high currents. Indeed, this information is already encompassed within the equilibrium potential and overpotential equations for spatially distributed models, where oxygen and hydrogen concentrations within the catalytic layers can be expressed. This is seen in this work equations \eqref{eq:equilibrium_voltage} and \eqref{eq:overpotential}. However, in most models, it remains necessary to retain $i_{lim}$ empirically because the current state of the art is not mature enough to take into account all the physical phenomena that impact voltage at high current densities.  Indeed, at high currents, liquid water emerges within the cell. This matter subsequently impacts the transport of oxygen and hydrogen to the triple point zones, making it more challenging. This results in a voltage drop \cite{lottinModellingOperationPolymer2009} for current densities lower than if there were no liquid water present. However, this has not been physically modeled in the existing literature, and $i_{lim}$ serves as an imperfect attempt to address this because it is detached from the physical variable that explains this phenomenon: the saturation in liquid water $\texttt{s}$.

Here, we propose a new coefficient, named limit liquid water saturation coefficient $\texttt{s}_{lim}$, which is indirectly added to the Butler-Volmer equation in \eqref{eq:overpotential} to physically consider the impact of catalyst layer flooding on its voltage. $U_{conc}$ is no longer useful. A physical interpretation of this coefficient will be proposed in the authors' future work. This proposition allows for a better connection between the equations and physics, which is valuable as it enables the observation, diagnosis and control of the factor responsible for the concentration drop: $\texttt{s}$. Additionally, this proposal easily links $\texttt{s}_{lim}$ to operating conditions in \eqref{eq:limit_liquid_saturation}, which is valuable for considering the stack beyond the arguable optimal conditions imposed by manufacturers. 

\begin{figure}[H]
	\centering
	\includegraphics[width=7cm]{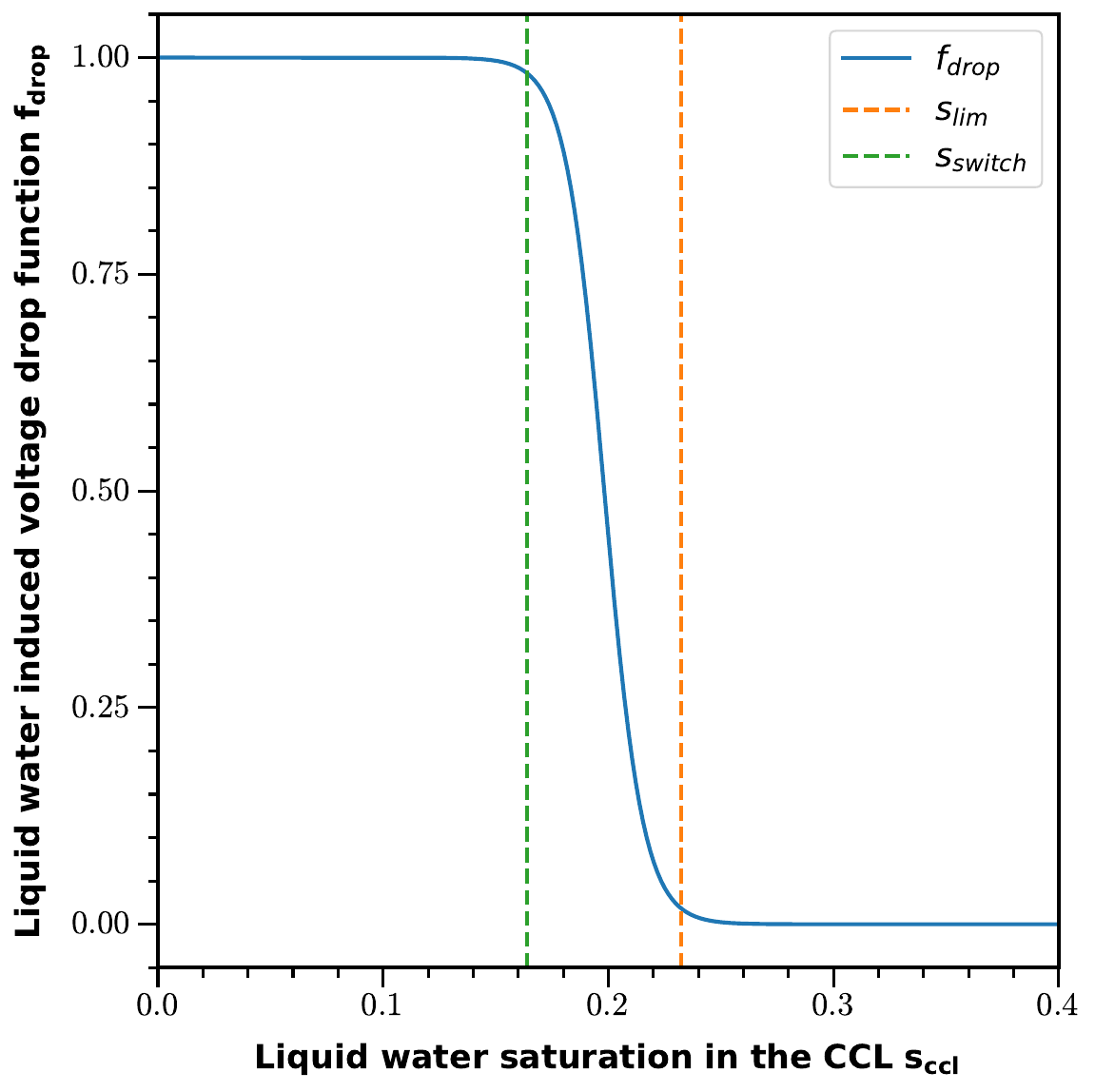}
	\caption{Plot of the liquid water induced voltage drop function, expressed as a function of the liquid water saturation in the CCL, \\ for $P_{des} = 2.5$ bar, $a_{s_{lim}} = 0.05$, $b_{s_{lim}} = 0.1075$ and $a_{switch} = 0.705$}
	\label{fig:f_drop}
\end{figure}

The proposed contribution here involves adding a new quantity to the Butler-Volmer equation: the liquid water induced voltage drop function $f_{drop}$. This function, expressed in \eqref{eq:f_drop} and shown in figure \ref{fig:f_drop}, equals to $0$ when the liquid water saturation of the cathodic catalytic layer $\texttt{s}_{ccl}$ exceeds the limit value of $\texttt{s}_{lim}$, resulting in an increase in overvoltage and, ultimately, a drop in voltage. When $\texttt{s}_{ccl}$ is sufficiently far from this limit, there is no impact of liquid water on the voltage, and therefore, $f_{drop}$ equals 1. In between, $f_{drop}$ strictly decreases towards 0. Indeed, experimentally, the concentration drop is not abrupt and extends over a few tenths of amperes per square centimeter. This is expressed by the fact that liquid water begins to significantly impact the voltage from a certain value of $\texttt{s}$, and this impact worsens with its increase until the stack stops. Thus, it is necessary to determine a boundary value for $\texttt{s}$ at which the voltage begins to drop, even before reaching $\texttt{s}_{lim}$. The authors propose considering $s_{switch}$, which takes a percentage of $\texttt{s}_{lim}$ as the boundary value for the start of voltage drop, as expressed in \eqref{eq:s_switch}. The proportionality coefficient $a_{switch}$ is an undetermined parameter of the model. Furthermore, $f_{drop}$ is built as a continuous and infinitely differentiable function, which is useful to avoid any fluctuations during numerical resolution.

Finally, given that $\texttt{s}$ is interpreted as a hindrance to the arrival of gases in the triple point areas, it is evident that $\texttt{s}_{lim}$ depends on the internal geometry of the stack materials, particularly the GDLs and CLs. Thus, its value inherently relies on the employed technology, making it impossible to establish a universal value. Even a slight modification in the porosity of the stack components would affect it. Therefore, it stands as a parameter specific to fuel cell design. Furthermore, it has been observed that $\texttt{s}_{lim}$ varies with the operating conditions. It is a linear function of the desired gas pressure set by the operator $P_{des}$, as demonstrated by the model validation section \ref{sec:validation_model}. Hence, its proposed expression in \eqref{eq:limit_liquid_saturation} involves $a_{s_{lim}}$ and $b_{s_{lim}}$ as two new undetermined parameters. The dependence of $\texttt{s}_{lim}$ on other operating conditions will be studied in future work. 
\section{Validation of the model's static behavior}
\label{sec:validation_model}
The developed model, including the matter flows, the voltage, and the auxiliaries, is implemented in Python. The corresponding programs are organized into a software package named AlphaPEM, which is freely accessible as open-source \cite{gassAlphaPEMOpensourceDynamic2025}. 

To validate the static behavior of the proposed model, a 1 kW EH-31 stack from EH Group \cite{zianeControleDiagnosticSans2022}, dated 2022, was utilized. The physical parameters of the stack, shown in table \ref{table:synthesis_accessible_physical_parameters}, were either measured in the laboratory or estimated based on conventional dimensions mentioned in the literature \cite{gassCriticalReviewProton2024}. Manufacturers seldom disclose these data; they typically provide only operating conditions. Subsequently, for this validation, experimental data on the same stack for different operational conditions are necessary. Here, polarization curves are employed as reference data. 
Among the operational conditions, it is the pressure within the stack (equal at the anode and cathode sides: $P_{a,des}$ = $P_{c,des}$ = $P_{des}$) that is altered, while other operational conditions remain constant. Their respective values are listed in table \ref{table:synthesis_manufacturer_operating_conditions}.

This validation only concerns the model's static behavior, since it solely relies on data representing the static states of the stack. To assess the dynamism of the model, forthcoming experiments will incorporate electrochemical impedance spectroscopy (EIS) curves. Additionally, this validation remains partial due to its utilization of operating conditions that do not fully capture the diversity of potential states within the stack, as discussed in section \ref{subsec:limits_model}.

\begin{table}[H]
	\centering
	\small
	\setlength{\abovedisplayskip}{-4pt} 
	\setlength{\belowdisplayskip}{-4pt} 
	\begin{tabularx}{13cm}{|YYYY|Y|} \hline
		
		\textbf{Symbol}
		& \multicolumn{2}{|>{\hsize=\dimexpr2\hsize+2\tabcolsep+\arrayrulewidth\relax}Y|}{
			\textbf{Accessible physical parameter}} 
		& \textbf{Measured value}
		& \textbf{Estimated value} \\ \hline
		
		$A_{act}$
		& \multicolumn{2}{|>{\hsize=\dimexpr2\hsize+2\tabcolsep+\arrayrulewidth\relax}Y|}{
			Active area $(m^{2})$} 
		& $8.5 \cdot 10^{-3}$ 
		& / \\
		
		$H_{mem}$
		& \multicolumn{2}{|>{\hsize=\dimexpr2\hsize+2\tabcolsep+\arrayrulewidth\relax}Y|}{
			Membrane thickness $(m)$} 
		& /
		& $2 \cdot 10^{-5}$  \\
		
		$H_{cl}$
		& \multicolumn{2}{|>{\hsize=\dimexpr2\hsize+2\tabcolsep+\arrayrulewidth\relax}Y|}{
			Catalyst layer thickness$(m)$} 
		& /
		& $10^{-5}$  \\
		
		$H_{gdl}$
		& \multicolumn{2}{|>{\hsize=\dimexpr2\hsize+2\tabcolsep+\arrayrulewidth\relax}Y|}{
			Gas diffusion layer thickness $(m)$} 
		& /
		& $2 \cdot 10^{-4}$  \\
		
		$H_{gc}$
		& \multicolumn{2}{|>{\hsize=\dimexpr2\hsize+2\tabcolsep+\arrayrulewidth\relax}Y|}{
			Gas channel thickness $(m)$} 
		& /
		& $5 \cdot 10^{-4}$  \\
		
		$W_{gc}$
		& \multicolumn{2}{|>{\hsize=\dimexpr2\hsize+2\tabcolsep+\arrayrulewidth\relax}Y|}{
			Gas channel width $(m)$} 
		& $4.5 \cdot 10^{-4}$ 
		& / \\
		
		$L_{gc}$
		& \multicolumn{2}{|>{\hsize=\dimexpr2\hsize+2\tabcolsep+\arrayrulewidth\relax}Y|}{
			Gas channel cumulated length $(m)$} 
		& $9.67$ 
		& /  \\ \hline
		
	\end{tabularx}
	\caption{Synthesis of the accessible physical parameters for the experimental fuel cell}
	\label{table:synthesis_accessible_physical_parameters}
\end{table}

\begin{table}[H]
	\centering
	\small
	\setlength{\abovedisplayskip}{-4pt} 
	\setlength{\belowdisplayskip}{-4pt} 
	\begin{tabularx}{13cm}{|YYYY|} \hline
		
		\textbf{Symbol}
		& \multicolumn{2}{|>{\hsize=\dimexpr2\hsize+2\tabcolsep+\arrayrulewidth\relax}Y|}{
			\textbf{Manufacturer operating conditions}} 
		& \textbf{Value} \\ \hline
		
		$T_{fc}$
		& \multicolumn{2}{|>{\hsize=\dimexpr2\hsize+2\tabcolsep+\arrayrulewidth\relax}Y|}{
			Cell temperature $(K)$} 
		& $347.15$ \\
		
		$P_{des}$
		& \multicolumn{2}{|>{\hsize=\dimexpr2\hsize+2\tabcolsep+\arrayrulewidth\relax}Y|}{
			Desired cell pressure (4 scenarios) $(bar)$} 
		& $1.5$ / $2.0$ / $2.25$ / $2.5$  \\
		
		$S_{a}$ / $S_{c}$
		& \multicolumn{2}{|>{\hsize=\dimexpr2\hsize+2\tabcolsep+\arrayrulewidth\relax}Y|}{
			Stoichiometries (anode/cathode)} 
		& $1.2$ / $2.0$  \\
		
		$\Phi_{a,des}$ / $\Phi_{c,des}$
		& \multicolumn{2}{|>{\hsize=\dimexpr2\hsize+2\tabcolsep+\arrayrulewidth\relax}Y|}{
			Desired entrance humidities (anode/cathode)} 
		& $0.4$ / $0.6$  \\ \hline
		
	\end{tabularx}
	\caption{Synthesis of the manufacturer operating conditions for the EH-31 experimental fuel cell}
	\label{table:synthesis_manufacturer_operating_conditions}
\end{table}

\begin{figure}[H]
	\centering
	\includegraphics[width=7cm]{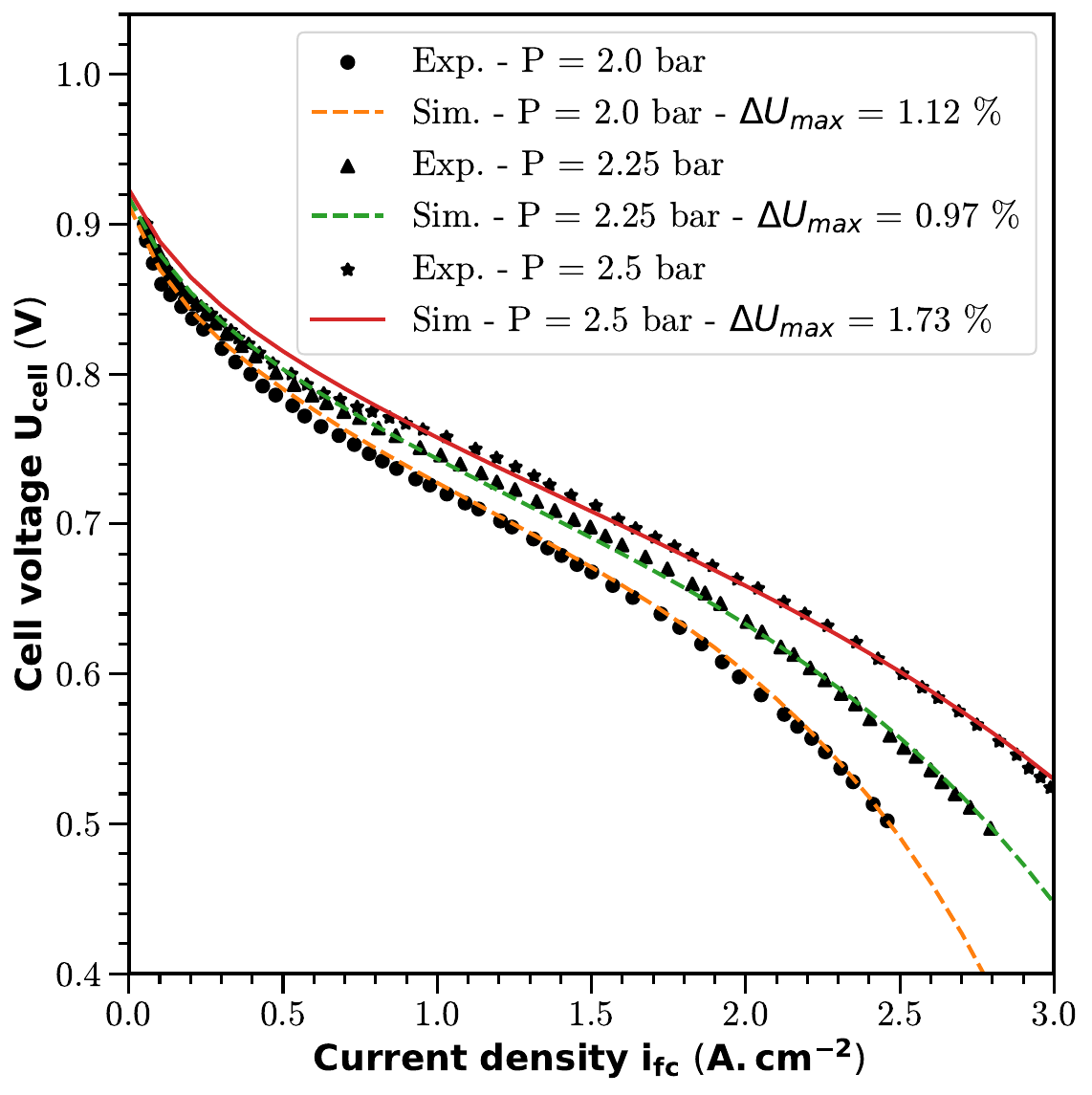}
	\caption{Comparison of polarization curves between simulation and experiment to validate the model's static behavior.}
	\label{fig:comparison_pola_curves_validation}
\end{figure}

The model validation process begins by calibrating the undetermined parameters. This calibration involves utilizing two sets of experimental polarization curves derived from the same cell but under distinct operating conditions. These sets serve as a reference for fine-tuning these parameters until achieving convergence between the model's results and the observed experimental outcomes. Here, the maximum voltage deviations $\Delta U_{max}$ between the model and experimental curves are below $1.2$ $\%$, indicating an excellent calibration. These curves are depicted in figure \ref{fig:comparison_pola_curves_validation} (the two dashed curves, at 2.0 and 2.25 bar), and their corresponding calibrated values are provided in table \ref{table:synthesis_calibrated_parameters}. Subsequently, the second validation step involves comparing the model outcomes with new experimental data obtained from the same cell, without altering any of the calibrated parameters, under varying operating conditions.  It is also noted that the tested data is under operating pressure outside the pressure range used for calibrating the model parameters. The model overfitting can therefore be excluded in the validation phase. Similarly, the maximum voltage deviation $\Delta U_{max}$ between the model and experimental curves is low, below $1.8$ $\%$. This result is shown in figure \ref{fig:comparison_pola_curves_validation} (the solid line curve at 2.5 bar). Hence, the model's static behavior has been validated through experimentation.

\begin{table}[H]
	\centering
	\small
	\setlength{\abovedisplayskip}{-4pt} 
	\setlength{\belowdisplayskip}{-4pt} 
	\begin{tabularx}{14.5cm}{|YYYY|} \hline
		
		\textbf{Symbol}
		& \multicolumn{2}{|>{\hsize=\dimexpr2\hsize+2\tabcolsep+\arrayrulewidth\relax}Y|}{
			\textbf{Undetermined physical parameters}} 
		& \textbf{Calibrated value} \\ \hline
		
		$i_{0,c}^{ref}$
		& \multicolumn{2}{|>{\hsize=\dimexpr2\hsize+2\tabcolsep+\arrayrulewidth\relax}Y|}{
			Referenced cathode exchange current density $(A.m^{-2})$} 
		& $2.79$ \\
		
		$\kappa_{co}$
		& \multicolumn{2}{|>{\hsize=\dimexpr2\hsize+2\tabcolsep+\arrayrulewidth\relax}Y|}{
			Crossover correction coefficient $(mol.m^{-1}.s^{-1}.Pa^{-1})$} 
		& $27.2$ \\
		
		$\kappa_{c}$
		& \multicolumn{2}{|>{\hsize=\dimexpr2\hsize+2\tabcolsep+\arrayrulewidth\relax}Y|}{
			Overpotential correction exponent} 
		& $1.61$ \\
		
		$\tau$
		& \multicolumn{2}{|>{\hsize=\dimexpr2\hsize+2\tabcolsep+\arrayrulewidth\relax}Y|}{
			Pore structure coefficient} 
		& $1.02$ \\
		
		$\varepsilon_{mc}$
		& \multicolumn{2}{|>{\hsize=\dimexpr2\hsize+2\tabcolsep+\arrayrulewidth\relax}Y|}{
			volume fraction of ionomer in the CLs} 
		& $0.399$ \\
		
		$R_{e}$
		& \multicolumn{2}{|>{\hsize=\dimexpr2\hsize+2\tabcolsep+\arrayrulewidth\relax}Y|}{
			Electron conduction resistance $(\Omega.m^{2})$} 
		& $5.70 \cdot 10^{-7}$ \\
		
		$\texttt{e}$
		& \multicolumn{2}{|>{\hsize=\dimexpr2\hsize+2\tabcolsep+\arrayrulewidth\relax}Y|}{
			Capillary exponent} 
		& $5$ \\
		
		$\varepsilon_{c}$
		& \multicolumn{2}{|>{\hsize=\dimexpr2\hsize+2\tabcolsep+\arrayrulewidth\relax}Y|}{
			GDL compression ratio} 
		& $0.271$ \\
		
		$\varepsilon_{gdl}$
		& \multicolumn{2}{|>{\hsize=\dimexpr2\hsize+2\tabcolsep+\arrayrulewidth\relax}Y|}{
			GDL porosity} 
		& $0.701$ \\
		
		$a_{s_{lim}}$, $b_{s_{lim}}$, $a_{switch}$
		& \multicolumn{2}{|>{\hsize=\dimexpr2\hsize+2\tabcolsep+\arrayrulewidth\relax}Y|}{
			$s_{lim}$ coefficients ($bar^{-1}$, $\emptyset$, $\emptyset$)} 
		& $0.0555$, $0.1051$, $0.63654$ \\  \hline
		
	\end{tabularx}
	\caption{Synthesis of the calibrated undetermined parameters for the EH-31 experimental fuel cell}
	\label{table:synthesis_calibrated_parameters}
\end{table}
\section{Results analysis}
\label{sec:results_model}

\subsection{Tracking internal states}

Under an arbitrary dynamic operating condition, the developed model enables monitoring within a cell of the water evolution, whether in the form of vapor, liquid, or dissolved matter in the membrane, characterized respectively by the variables $C_v$, $\texttt{s}$, or $\lambda$. It also tracks the evolution of dihydrogen, dioxygen, and nitrogen, characterized by the variables $C_{H_2}$, $C_{O_2}$, and $C_{N_2}$. These variables are evaluated at several nodes within the cell, and the variables with indices agdl or cgdl refer to the node in the center of the corresponding GDL. Additionally, data regarding matter flows between these nodes ($J$) are also accessible. Furthermore, the evolution of pressures $P$ and humidities $\Phi$ within the auxiliary manifolds can also be tracked. Finally, the cell voltage over time $U_{cell}$ is calculated from these internal states.

\begin{figure}[H]
	\centering
	\begin{subfigure}[b]{0.49\textwidth}
		\centering
		\includegraphics[width=\textwidth]{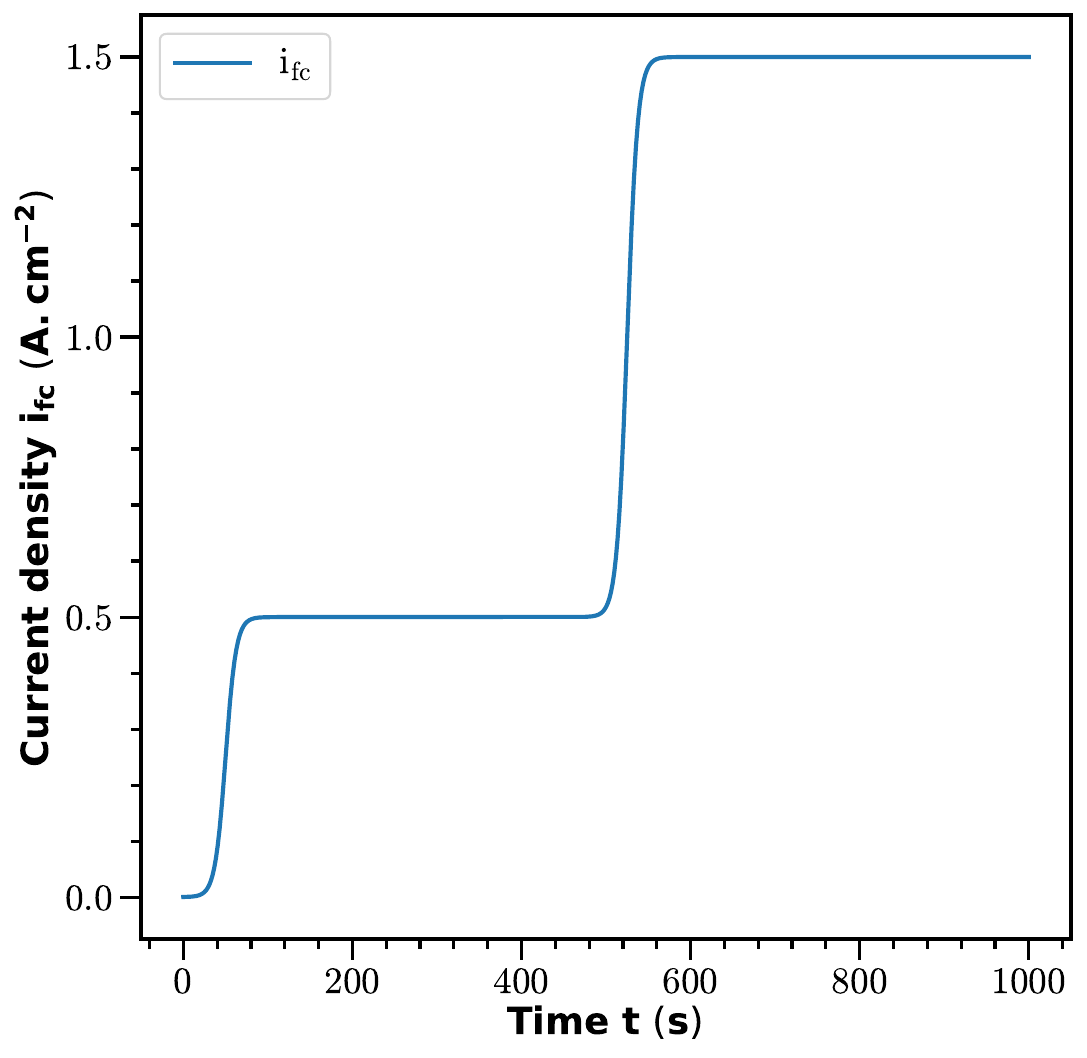}
		\caption{Current density evolution}
		\label{fig:step_current_syn_ifc}
	\end{subfigure}
	\hfill
	\begin{subfigure}[b]{0.496\textwidth}
		\centering
		\includegraphics[width=\textwidth]{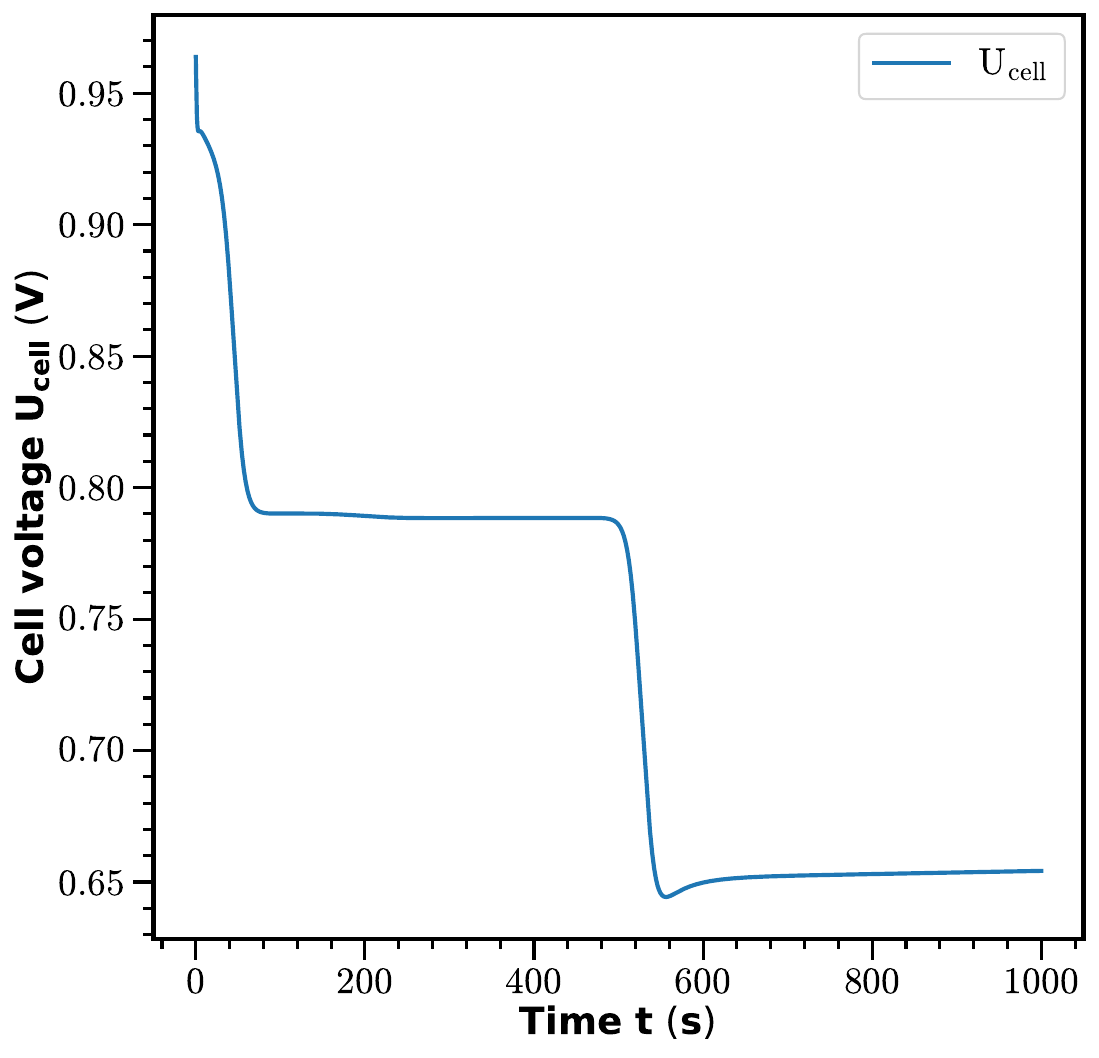}
		\caption{Voltage evolution}
		\label{fig:step_current_syn_Ucell}
	\end{subfigure}
	\hfill
	\begin{subfigure}[b]{0.49\textwidth}
		\centering
		\includegraphics[width=\textwidth]{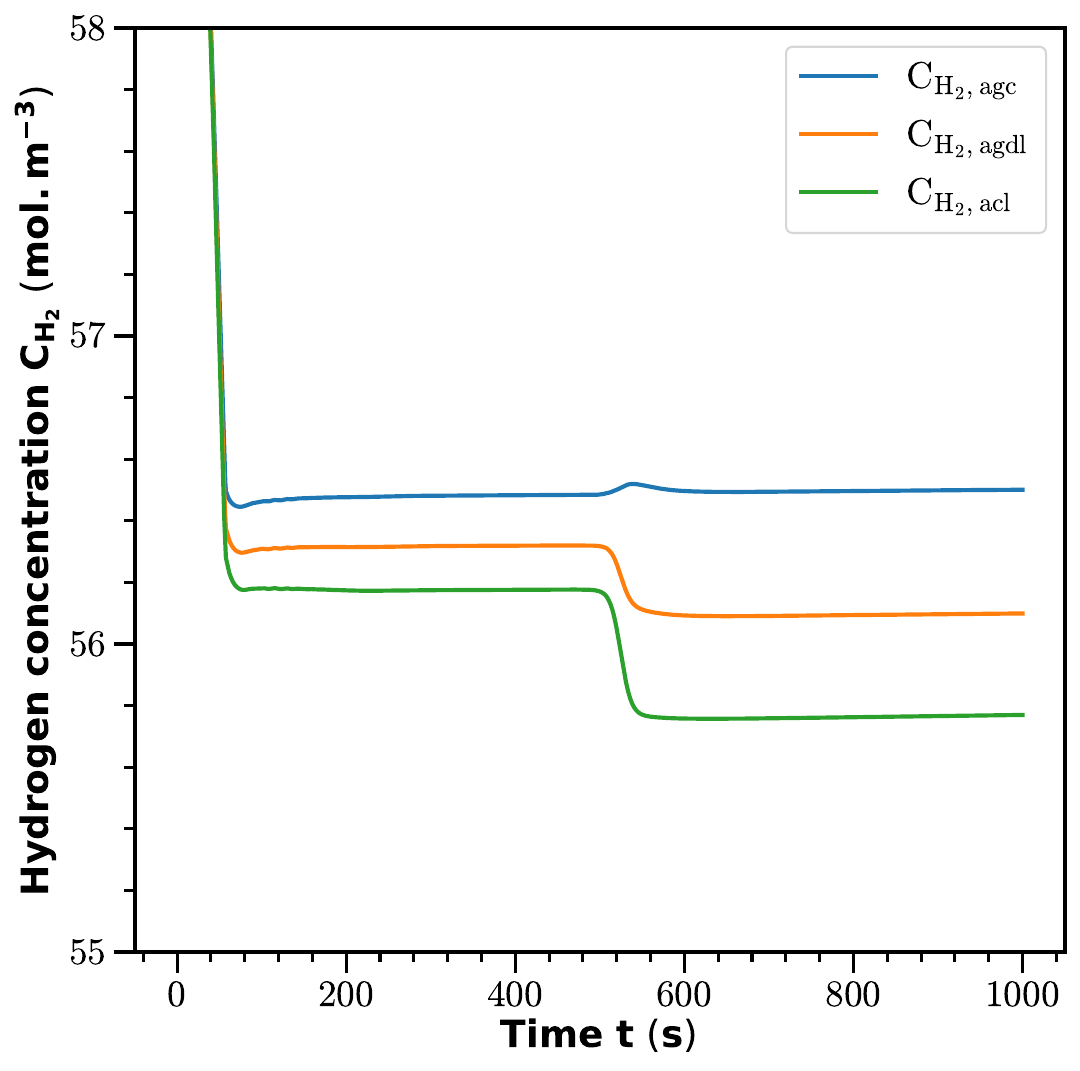}
		\caption{Hydrogen evolution}
		\label{fig:step_current_syn_C_H2}
	\end{subfigure}
	\hfill
	\begin{subfigure}[b]{0.49\textwidth}
		\centering
		\includegraphics[width=\textwidth]{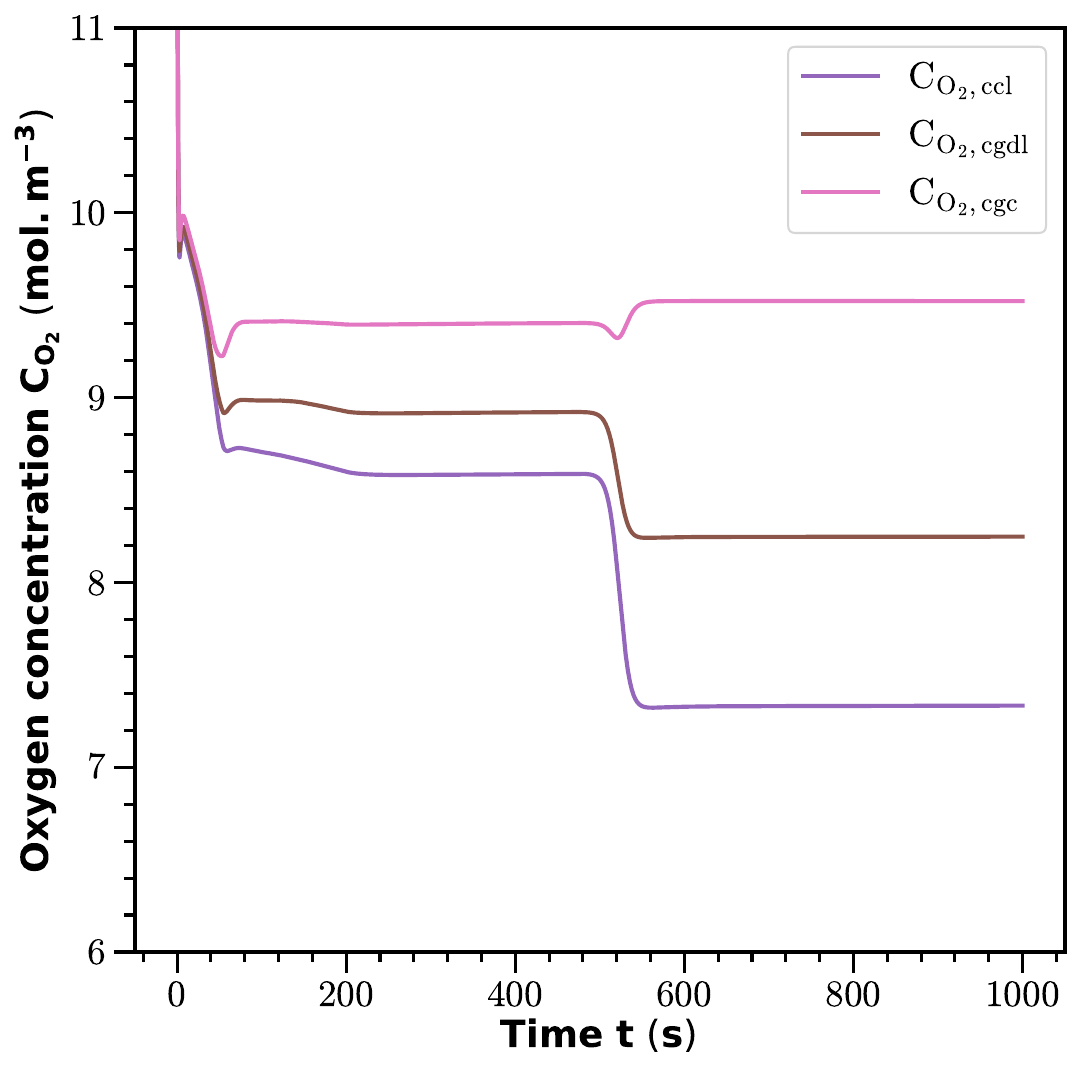}
		\caption{Oxygen evolution}
		\label{fig:step_current_syn_C_O2}
	\end{subfigure}
	\caption{Internal states of a PEM fuel cell system for two current density steps, computed by AlphaPEM (1/3).}
	\label{fig:step_current_syn_1}
\end{figure}

\begin{figure}[H]
	\centering
	\begin{subfigure}[b]{0.493\textwidth}
		\centering
		\includegraphics[width=\textwidth]{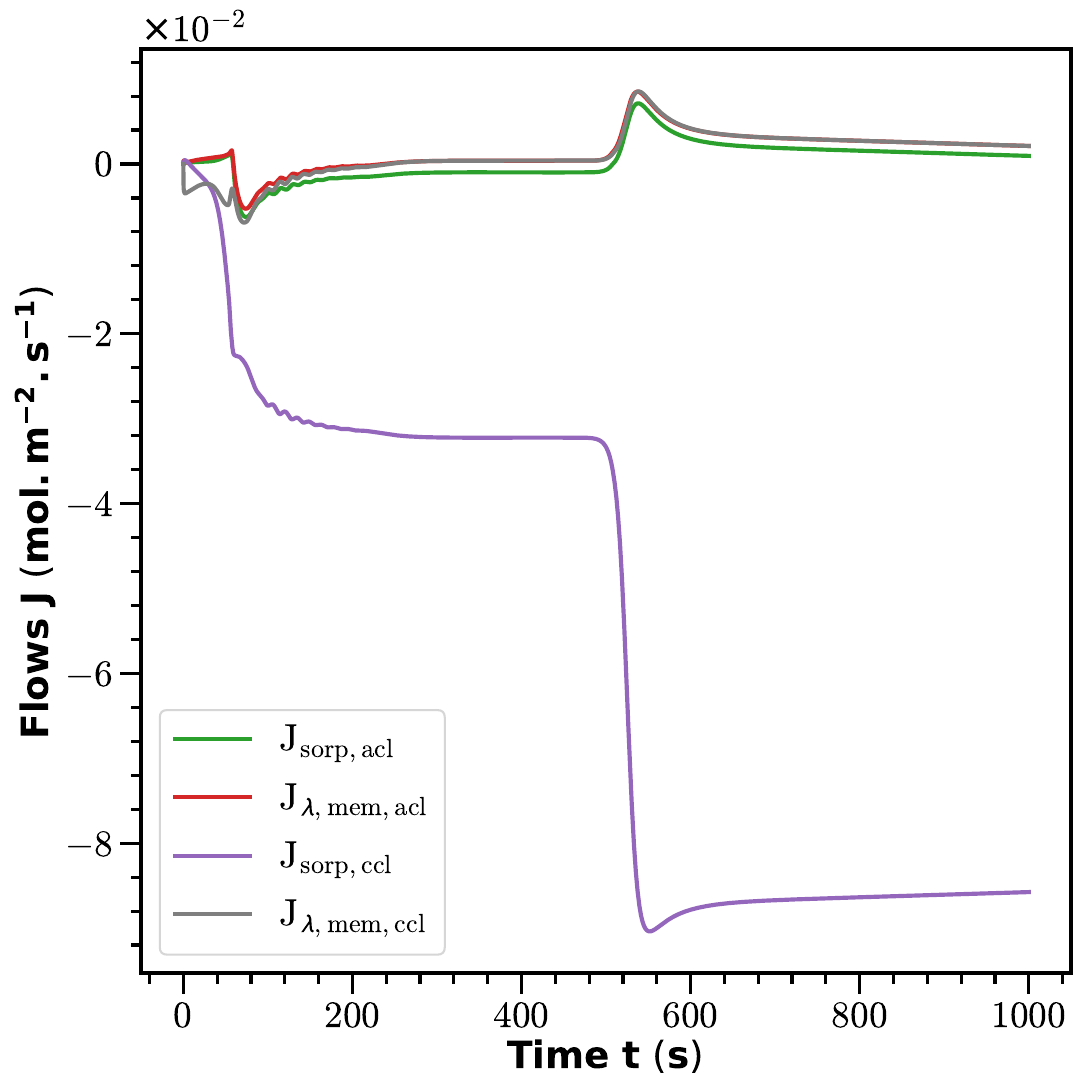}
		\caption{Flows evolution}
		\label{fig:step_current_syn_J}
	\end{subfigure}
	\hfill
	\begin{subfigure}[b]{0.49\textwidth}
		\centering
		\includegraphics[width=\textwidth]{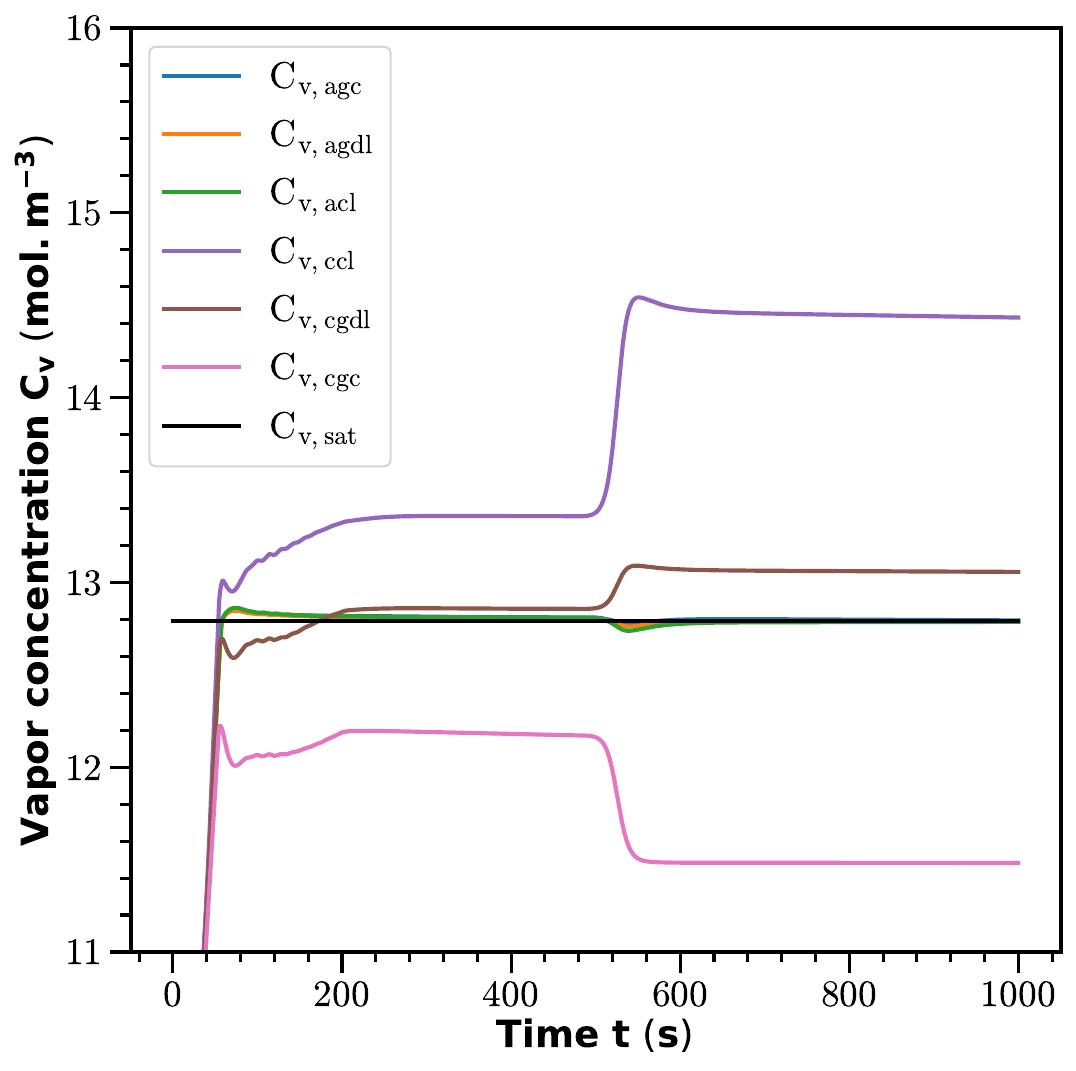}
		\caption{Vapor evolution}
		\label{fig:step_current_syn_Cv}
	\end{subfigure}
	\hfill
	\begin{subfigure}[b]{0.505\textwidth}
		\centering
		\includegraphics[width=\textwidth]{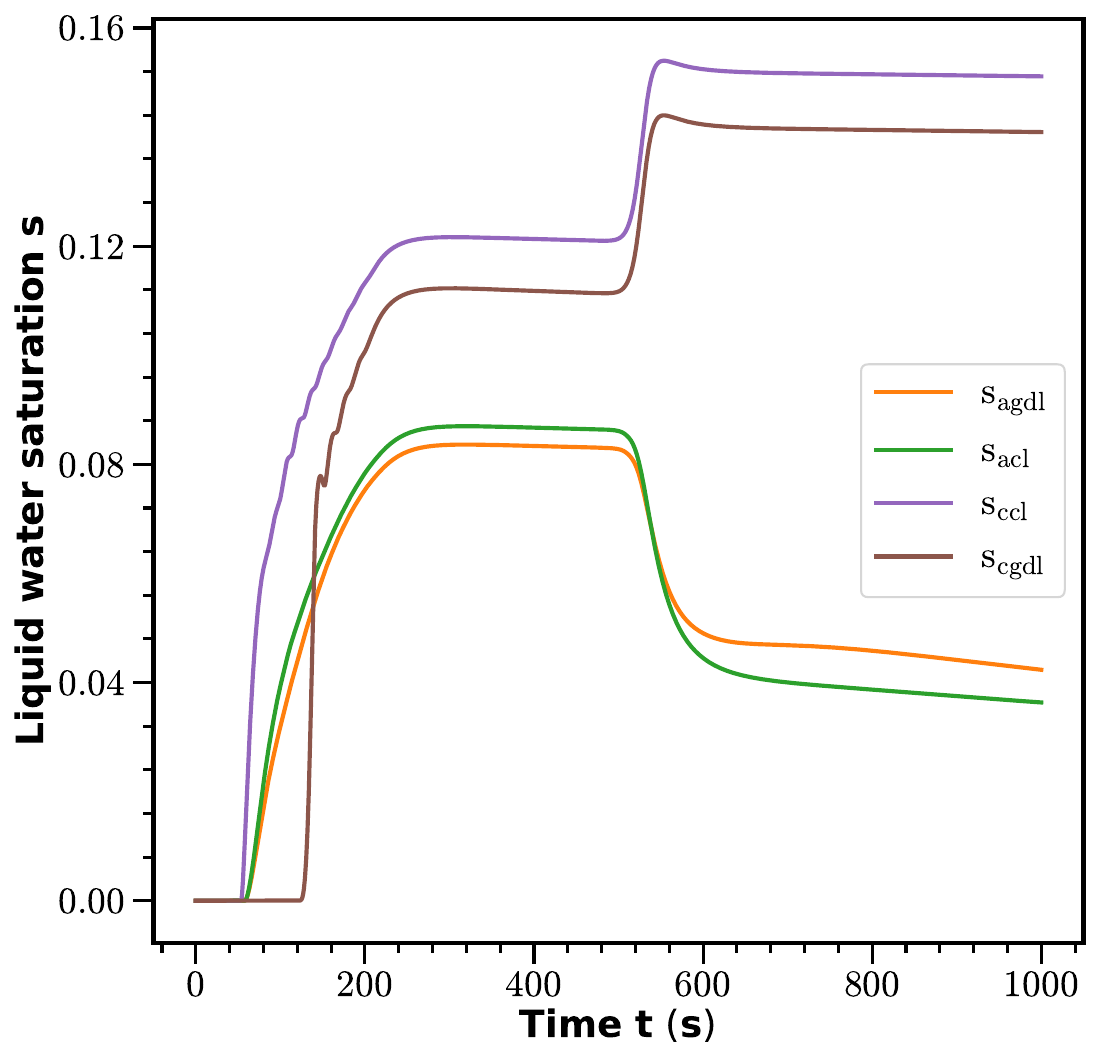}
		\caption{Liquid water evolution}
		\label{fig:step_current_syn_s}
	\end{subfigure}
	\hfill
	\begin{subfigure}[b]{0.49\textwidth}
		\centering
		\includegraphics[width=\textwidth]{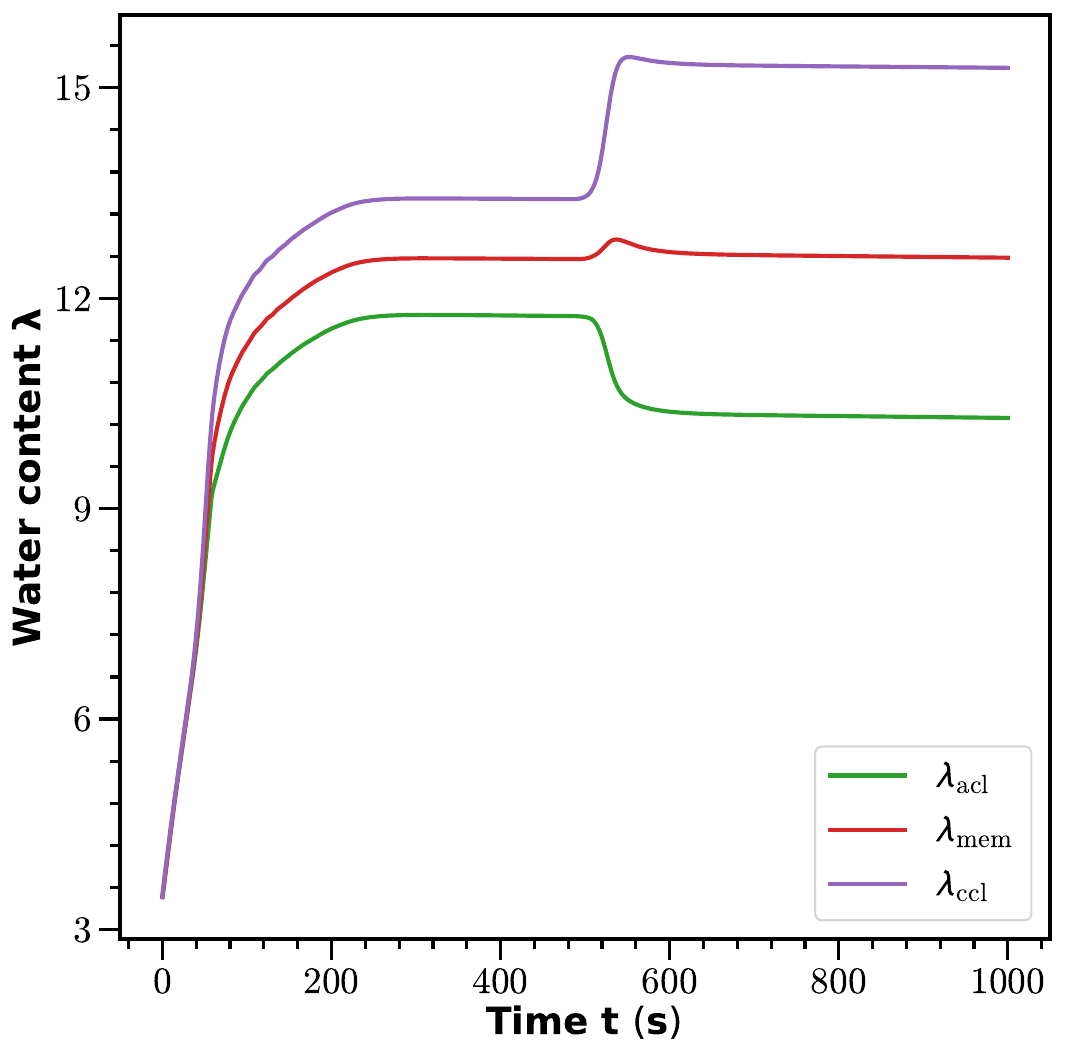}
		\caption{Dissolved water evolution}
		\label{fig:step_current_syn_lambda}
	\end{subfigure}
	\caption{Internal states of a PEM fuel cell system for two current density steps, computed by AlphaPEM (2/3).}
	\label{fig:step_current_syn_2}
\end{figure}

\begin{figure}[H]
	\centering
	\begin{subfigure}[b]{0.508\textwidth}
		\centering
		\includegraphics[width=\textwidth]{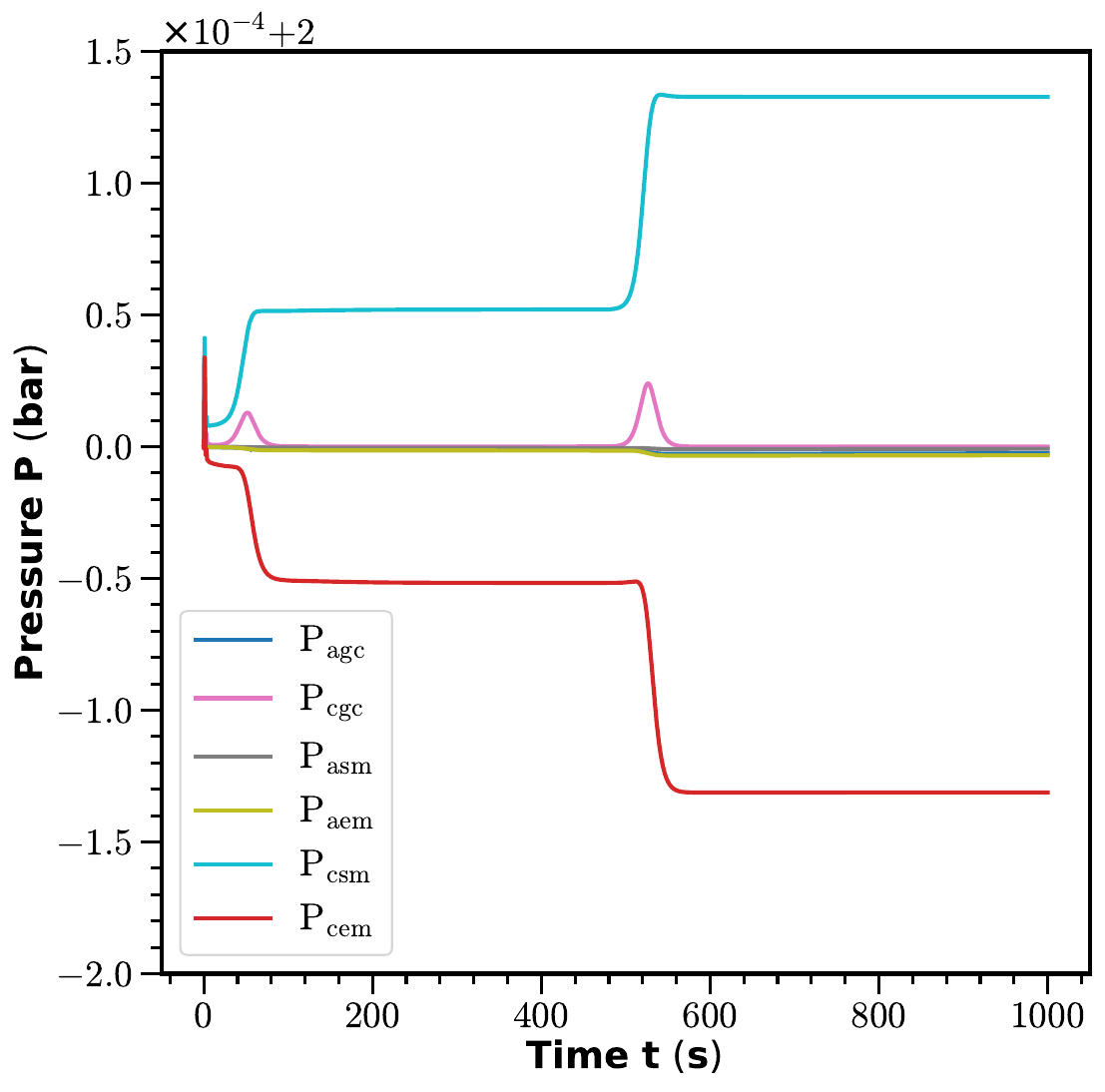}
		\caption{Pressures evolution}
		\label{fig:step_current_syn_P}
	\end{subfigure}
	\hfill
	\begin{subfigure}[b]{0.485\textwidth}
		\centering
		\includegraphics[width=\textwidth]{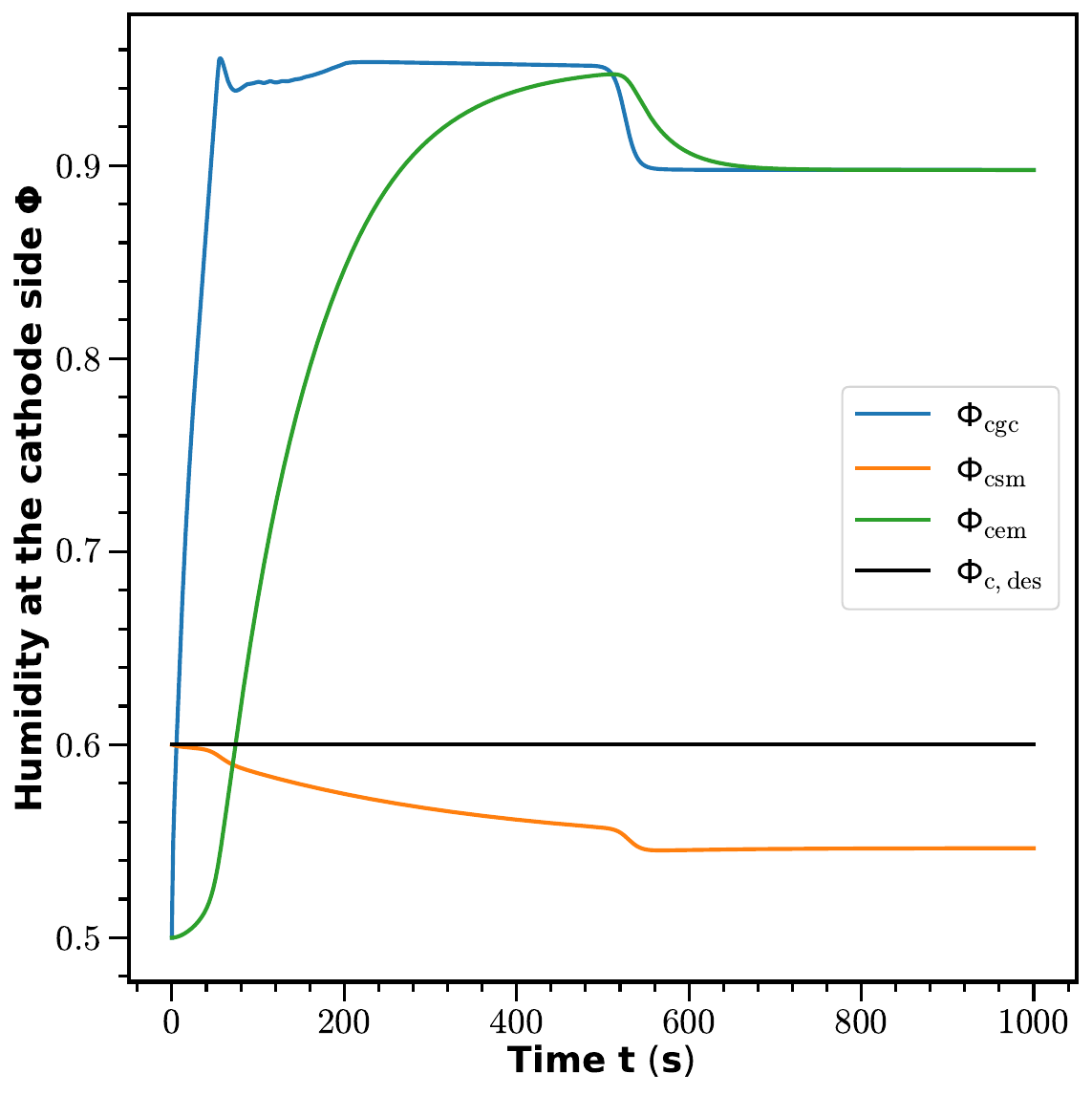}
		\caption{Humidity evolution at the cathode side}
		\label{fig:step_current_syn_Phi_c}
	\end{subfigure}
	\caption{Internal states of a PEM fuel cell system for two current density steps, computed by AlphaPEM (3/3).}
	\label{fig:step_current_syn_3}
\end{figure}

Several results of the calibrated model are shown in figures \ref{fig:step_current_syn_1}, \ref{fig:step_current_syn_2} and \ref{fig:step_current_syn_3}, under pressure $P_{des} = 2.0$ bar. In this study case, a step-shape current density is applied, ranging from $0$ $A.cm^{-2}$ to $0.5$ $A.cm^{-2}$ at the start of the experiment, and then from $0.5$ $A.cm^{-2}$ to $1.5$ $A.cm^{-2}$ at $500s$, as seen in figure \ref{fig:step_current_syn_ifc}. The variables are initialized to the values they would have in steady-state conditions, with zero current density, if they were subjected to the pressure, humidity, and temperature of the desired operating conditions. For simplicity, it is assumed that the variables within each cell are initially subject to the average of the anodic and cathodic pressures and humidity. The experiment virtually lasts $1000s$.

The advantage of performing a double step-shape current density is to eliminate the need for initial condition values in the analysis of the results. Indeed, the first step-shape current density allows the system to reach a steady state within the fuel cell, with a waiting time of 500 seconds. Since the fuel cell is controlled, it will always reach the same steady state regardless of its initial conditions, which only influence its transition to this stationary state. Thus, after around 500 seconds, the fuel cell operates in a state with realistic internal values. The second step-shape current density can then be studied under standardized conditions.

The results generally follow the expected pattern within the cell: an increase in current density leads to increased flows, reduced reactants, and increased water content. However, it is necessary to further examine certain variables to clarify their behavior.
Firstly, the reactants in the bipolar plates, characterized by $C_{H_2,agc}$ and $C_{O_2,cgc}$ figures \ref{fig:step_current_syn_C_H2} and \ref{fig:step_current_syn_C_O2}, do not exhibit significant changes and tend to slightly increase, unlike the reactants in the membrane electrode assembly (MEA) $C_{H_2,agdl}$, $C_{H_2,acl}$, $C_{O_2,cgdl}$ and $C_{O_2,ccl}$. This can be explained by the fact that $C_{H_2,agc}$ and $C_{O_2,cgc}$ are less sensitive to the chemical activity within the MEA, as the stack is designed to stabilize the pressure within the bipolar plates using a backpressure valve. The slight fluctuations are attributed to changes in the composition of this gas mixture, with a decrease in vapor concentration ($C_{v,agc}$ and $C_{v,cgc}$ figure \ref{fig:step_current_syn_Cv}) occurring at high currents due to its expulsion by the increased gas flow rates involved.

\begin{figure}[H]
	\centering
	\includegraphics[width=7cm]{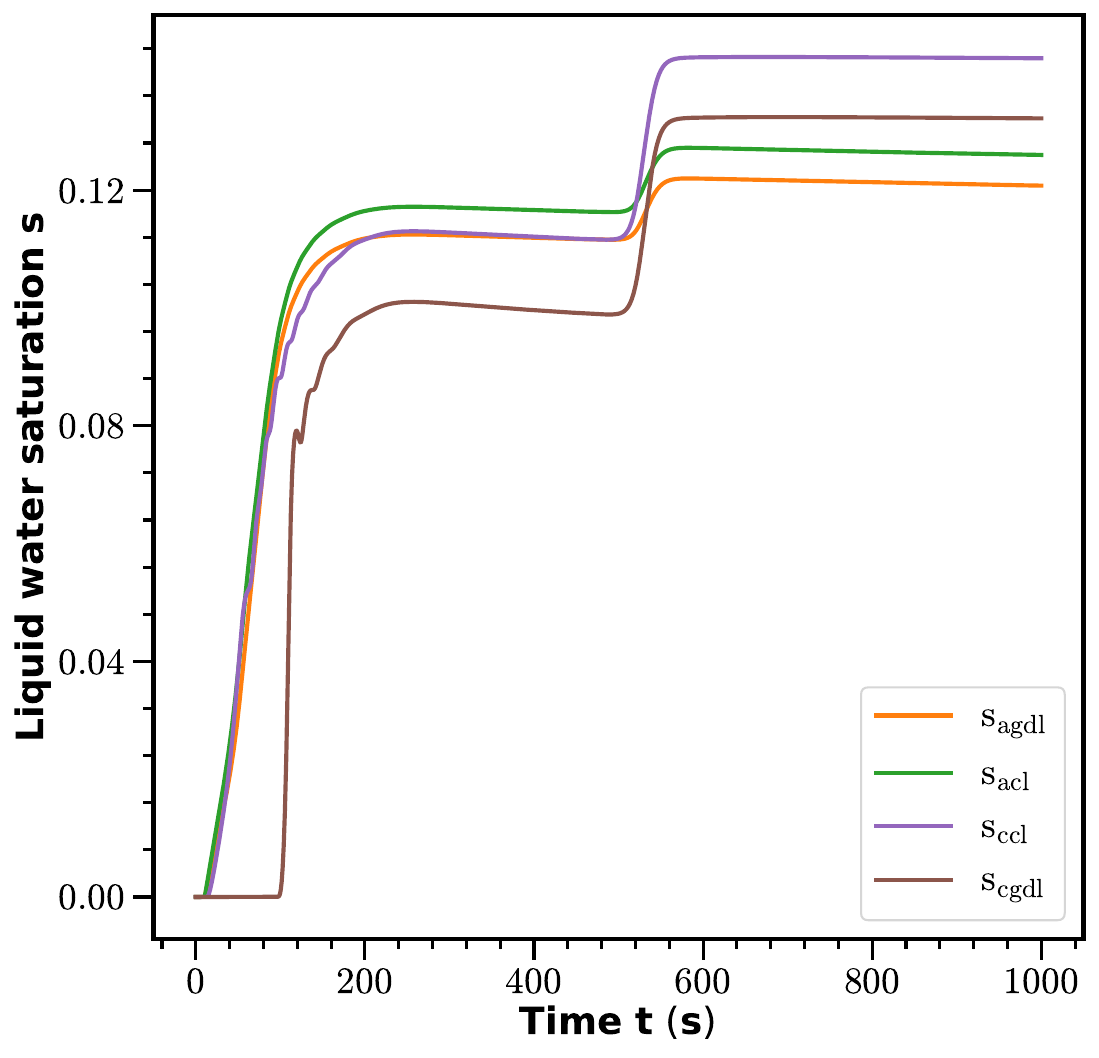}
	\caption{Evolution of liquid water within the cell for a membrane and a catalytic layer three times thinner.}
	\label{fig:step_current_s}
\end{figure}

Then, it is surprising that the behavior of water at the anode differs from that at the cathode, regardless of its form (vapor with $C_{v,agdl}$ and $C_{v,acl}$ figure \ref{fig:step_current_syn_Cv}, liquid with $\texttt{s}_{agdl}$ and $\texttt{s}_{acl}$ figure \ref{fig:step_current_syn_s}, or dissolved with $\lambda_{agdl}$ and $\lambda_{acl}$ figure \ref{fig:step_current_syn_lambda}): it decreases with current density (except at low currents < $0.5$ $A.cm^{-2}$ where it increases with current density, even after leaving the initial state). This can be explained by the existence of two opposing phenomena. On one hand, more water is created at the cathode with increasing current and passes through the membrane towards the anode. On the other hand, the flow of gases circulating in the bipolar plates also increases, making it easier to remove water from the MEA. As these flows are of the same order of magnitude, it is not easy to predict the evolution of water vapor in the anode. This depends on several parameters, such as the stoichiometry and geometric parameters like the thicknesses of the membrane and the thicknesses of the MEA. To illustrate this point, the same experiment was repeated with a threefold reduction in the thickness of the membrane and the catalytic layer, significantly reducing the resistance of the membrane to the passage of water from the cathode to the anode. Thus, the decrease in liquid water at the anode side is no longer visible and has been replaced by an increase, as shown in Figure \ref{fig:step_current_s}.

Furthermore, the impact of auxiliary dynamics is particularly evident in the evolution of oxygen concentrations with $C_{O_2,cgc}$ figure \ref{fig:step_current_syn_C_O2}, or equivalently $P_{cgc}$ figure \ref{fig:step_current_syn_P} (which influences $C_{O_2,cgdl}$ and $C_{O_2,ccl}$), leading to fluctuations in concentrations with each change in current density. This phenomenon does not occur when the cell is modeled without auxiliaries. However, in this model, the other variables are less affected than $C_{O_2,cgc}$ by the presence of auxiliaries.

However, there is a fluctuation in most internal states when a current density step is crossed, especially concerning water (see $C_{v,ccl}$ figure \ref{fig:step_current_syn_Cv} for example). It is characterized by a slight overshoot in the equilibrium value. This can be explained by the sudden increase in current that causes a sudden production of water in the cell. The discharge of this water is not sudden and possesses some inertia, leading to a transient over-accumulation of matter, namely a peak. This observed dynamic phenomenon is of interest, considering that the amount of water can affect the cell's voltage and potentially damage it. Thus, in energy management strategies, it might be interesting to slow down this increase in current density attributed to the fuel cell by temporarily compensating the energy demand with other electricity sources, such as batteries. Consequently, these observed peaks will disappear.

Next, liquid water saturation sometimes evolves with slight fluctuations, notably observed figure \ref{fig:step_current_syn_s} around $200 s$ for $\texttt{s}_{ccl}$ and $\texttt{s}_{cgdl}$. These fluctuations subsequently impact other variables, such as $C_{v,cgdl}$, $C_{v,cgc}$, $S_{sorp,acl}$, $J_{\lambda,mem,acl}$, $S_{sorp,ccl}$, and $J_{\lambda,mem,ccl}$. These are minor numerical errors resulting from an insufficiently high number of nodes in each GDL, as discussed in section \ref{subsubsec:nodal_model}. Here, it was chosen not to use an excessively high number of nodes to avoid significantly increasing computation times, even at the cost of a slight loss in precision in the results. Indeed, quadrupling $n_{gdl}$ is necessary to achieve nearly perfectly smooth results, which triples the computation times.

It is also noteworthy to observe that water vapor concentrations $C_v$ can exceed the saturation vapor value $C_{v,sat}$ figure \ref{fig:step_current_syn_Cv}. This can be explained by the dynamic equilibrium at stake. On one hand, surpassing the water vapor saturation threshold triggers the condensation of this vapor into liquid water. However, this condensation is not instantaneous and depends on a time constant $\gamma_{cond}$ embedded within the model. On the other hand, the stack continues to produce large amounts of water that feed into the water vapor. Indeed, in this model, it has been assumed that water production occurs necessarily in a dissolved manner. The current state of research does not allow us to determine in what form water appears immediately after the chemical redox reaction between hydrogen and oxygen \cite{jiaoWaterTransportPolymer2011, gassCriticalReviewProton2024}. A choice must therefore be made. Furthermore, in this model, the water flows between the membrane and the catalytic layer necessarily occur between a dissolved form and a vapor form. Only thereafter is condensation possible. Water production in the cell therefore directly involves vapor water supply. The supply flow of water vapor and condensation thus oppose each other, resulting in a dynamic equilibrium that can exceed the saturation vapor point, as long as the cell operates. If the time constant associated with condensation, $\gamma_{cond}$, is increased sufficiently, this phenomenon disappears, and $C_{sat}$ becomes the actual limit of the water vapor concentration. However, the value chosen for $\gamma_{cond}$ in the authors' model corresponds to that recommended by Hua Meng in a dedicated study \cite{mengTwoPhaseNonIsothermalMixedDomain2007}. Thus, this oversaturation phenomenon is acceptable.

Inside the auxiliaries, it is also remarkable to note that the pressure difference between the manifolds and the bipolar plates, shown figure \ref{fig:step_current_syn_P}, is low in this model, on the order of $1$ to $10 Pa$, which is not realistic. This stems, on the one hand, from the unmodeled pressure losses, and on the other hand, from the choice of equations \eqref{eq:Wa_sm_in}, \eqref{eq:Wa_sm_out}, \eqref{eq:Wa_em_in}, \eqref{eq:Wc_sm_out}, and \eqref{eq:Wc_em_in} which concern the incoming or outgoing matter flows from the manifolds and are based on simplifying assumptions. This is an aspect that needs improvement in the model.

Moreover, it is interesting to discuss the evolution of humidity in the auxiliaries, as shown in the curve \ref{fig:step_current_syn_Phi_c}. The supply manifold receives a controlled water flow which is at the desired humidity level, while also delivering a water flow to the cell. It stabilizes at a value lower than the desired humidity. This is a consequence of the chosen humidity control strategy, which focuses on the water flow entering the supply manifold rather than the humidity level within the supply manifold itself. Additionally, it can be observed that the humidity in the exhaust manifold stabilizes at the same level as the humidity in the gas channel. This humidity also corresponds to that of the flow exiting the cell, as the current model is one-dimensional.

\subsection{AlphaPEM computational efficiency}

This $1000 s$ simulation was conducted on a workstation featuring an Intel Core i9-11950H @ 2.60 GHz processor and required $17s$ of computation time. Simulating a polarization curve takes $9s$. Therefore, the model implemented within AlphaPEM operates within the same order of magnitude as other 1D simulators mentioned in the literature \cite{xuReduceddimensionDynamicModel2021}, is two orders of magnitude faster than a 1D model from commercial software like Comsol Multiphysics \cite{xuReduceddimensionDynamicModel2021}, and four to five orders of magnitude faster than 1D+1D, 3D+1D, or 3D models from the literature \cite{yangInvestigationPerformanceHeterogeneity2020, xieValidationMethodologyPEM2022, tardyInvestigationLiquidWater2022}. The computation times obtained by AlphaPEM are thus compatible with uses in embedded applications. It is important to note that while the model's computational speed is significantly enhanced, its precision is inherently lower compared to models simulating higher dimensional spaces.

\subsection{Limits of the model}
\label{subsec:limits_model}

\begin{figure}[H]
	\centering
	\includegraphics[width=7cm]{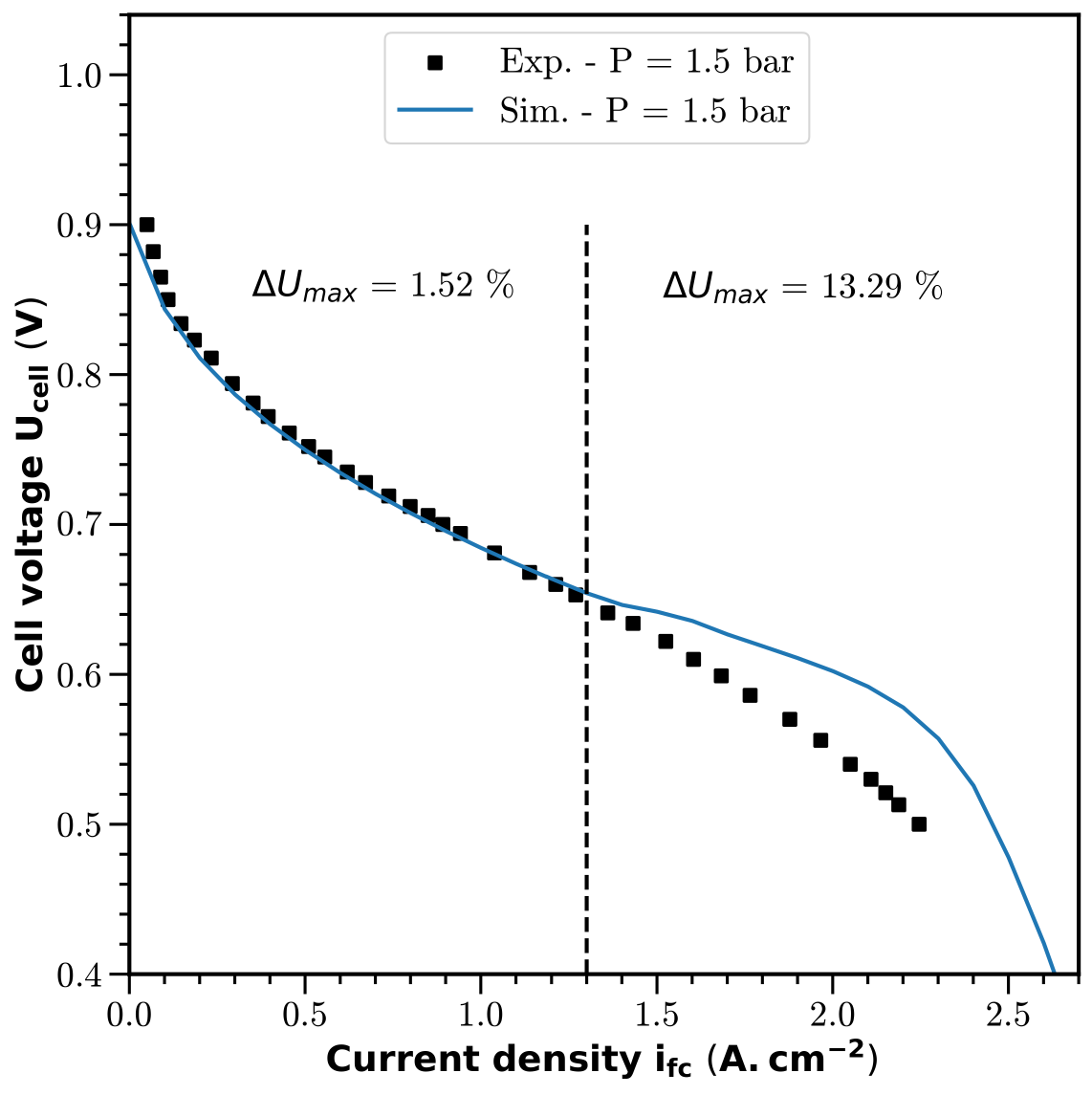}
	\caption{Comparison of polarization curves between simulation and experiment at 1.5 bar.}
	\label{fig:pola_issue}
\end{figure}

Despite the excellent agreements observed in section \ref{sec:validation_model} between the experimental and model results at pressures of 2.0, 2.25, and 2.5 bar, the comparison is less favorable at a lower pressure of 1.5 bar, as illustrated in figure \ref{fig:pola_issue}. Specifically, the error remains low for current densities below $1.3$ $A.cm^{-2}$, with $\Delta U_{max} = 1.5$ $\%$ within this range, but increases significantly for higher current densities.

This variation arises from the condensation of water within the cell, starting exactly from $i_{fc} = 1.3$ $A.cm^{-2}$, whereas liquid water was consistently present for all current density levels in previous experiments. It is plausible that the limited theoretical understanding of water sorption in catalytic layers, as criticized in the authors' prior work \cite{gassCriticalReviewProton2024}, causes inaccurate simulation of the transition from a humid gas without condensed water to a gas saturated with vapor and with liquid water. This could result in the significant errors observed in this simulated voltage. Thus, the authors urge the scientific community to enhance the theory describing the evolution of water in its various states within each cell. 

It is also conceivable that this deviation arises from the methodology employed by stack manufacturers in experimentally measuring the polarization curve. If the measurement is conducted dynamically rather than statically, it could impact the results. Dynamic measurement of the polarization curve involves initially balancing the stack at a nominal operating point, then rapidly sweeping through the entire current density range without achieving perfect equilibrium for each current density. Consequently, if the stack lacks adequate time for proper balancing, the amount of liquid water at a given current density might differ compared to a static scenario, resulting in disparities in the measured concentration drop. This effect could be particularly pronounced at a pressure of $1.5$ bar, as the nominal operating point might not initially yield liquid water, for instance, if sets at $1.0$ $A.cm^{-2}$, unlike in other considered scenarios. Hence, new polarization curve tests with assured equilibrium at each current density point would be imperative to verify this hypothesis.

Additionally, it is crucial to acknowledge the limited scope of validating a PEM fuel cell model solely based on three polarization curves. These curves, which vary only in pressure from 2.0 to 2.5 bar, fail to encompass the full range of physical scenarios occurring within one cell. Indeed, the transition between a humid gas without liquid water and a gas saturated with vapor containing liquid water is notably absent with these operating conditions. Consequently, the accuracy of the results is contingent upon specific conditions, rendering the model unreliable for all scenarios. It would be beneficial to develop a routine, under specified operating conditions, that ensures comprehensive coverage of all relevant physical phenomena within the cell, for its static validation with polarization curves.

\section{Conclusion}

Multi-physics models allow increasing the available information to better control PEM fuel cells, which is valuable considering the impossibility of placing sensors inside a cell. Currently, most existing models either provide a very detailed description of the internal states of the cell but require a very high computational cost, such as computational fluid dynamics models, or are fast but provide summary information about the cell, such as lumped-parameter models.

This work aims to find a better compromise to combine result accuracy and execution speed. Thus, a one-dimensional, dynamic, two-phase, isothermal, and finite-difference model of the PEMFC system has been developed, and its static behavior has been validated against several published experimental polarization curves. This model runs two orders of magnitude faster than 1D models from the commercial software Comsol Multiphysics and up to five orders of magnitude faster than 3D models from the literature. It remains compatible with embedded applications and provides more precision than lumped-parameter models.

In addition, a new coefficient has been introduced to replace the limit current density coefficient ($i_{lim}$). This coefficient, the limit liquid water saturation coefficient ($s_{lim}$), also determines the voltage drop at high current densities. $s_{lim}$ offers the added advantage of establishing a physical connection between this voltage drop, the internal states of the cell, and the operating conditions. Moreover, this parameter has been proven to be a function of the pressure imposed by the operators $P_{des}$. 

In upcoming researches, experimental verification will be conducted to determine whether $s_{lim}$ is dependent on other operating conditions, such as the temperature $T_{fc}$, and a physical interpretation of this coefficient will be proposed. Additionally, the model will undergo refinement through the incorporation of heat exchange modeling, the extension to a "1D+1D" model and the simulation of electrochemical impedance spectroscopy curves, all while maintaining computational efficiency. Further attention will be given to enhancing the control design of the model. Finally, the algorithm for this fuel cell model, released as open-source software and named AlphaPEM, will serve as a robust tool for future researchers needing a modifiable complex physics-based model for their investigations.

\section{Acknowledgments}

This work has been supported by French National Research Agency via project DEAL (Grant no. ANR-20-CE05-0016-01), the Region Provence-Alpes-Côte d'Azur, the EIPHI Graduate School (contract ANR-17-EURE-0002) and the Region Bourgogne Franche-Comté.

\renewcommand\nomgroup[1]{
	\item[\bfseries
	\ifstrequal{#1}{A}{Physical quantities}{
		\ifstrequal{#1}{B}{Mathematical symbols}{
			\ifstrequal{#1}{C}{Subscripts and superscripts}{
				\ifstrequal{#1}{D}{Abbreviation}{}}}}
	]}

\nomenclature[A,1]{$n$}{number of moles $( mol )$ }
\nomenclature[A,1]{$n_{gdl}$}{number of nodes inside each GDL}
\nomenclature[A,1]{$n_{cell}$}{number of cells inside the simulated stack}
\nomenclature[A,1]{$J$}{molar/mass transfer flow $(mol.m^{-2}.s^{-1} / kg.m^{-2}.s^{-1} )$}
\nomenclature[A,1]{$i$}{current density per unit of cell active area $( A.m^{-2} ) $}
\nomenclature[A,1]{$i_n$}{internal current density $( A.m^{-2} ) $}
\nomenclature[A,1]{$i_{lim}$}{limit current density coefficient}
\nomenclature[A,1]{$F$}{Faraday constant $( C.mol^{-1} ) $}
\nomenclature[A,1]{$M$}{molecular weight $( kg.mol^{-1} )$}
\nomenclature[A,1]{$T_{fc}$}{fuel cell temperature $( K )$}
\nomenclature[A,1]{$a_{w}$}{water activity in the pores of the CL}
\nomenclature[A,1]{$D$}{diffusion coefficient of water in the membrane $( m^{2}.s^{-1} )$}
\nomenclature[A,1]{$x$}{space variable $( m )$}
\nomenclature[A,1]{$P$}{pressure $(Pa)$}
\nomenclature[A,1]{$r_{f}$}{carbon fiber radius $(m)$}
\nomenclature[A,1]{$H$}{thickness $( m ) $}
\nomenclature[A,1]{$C$}{molar concentration $( mol.m^{-3} )$}
\nomenclature[A,1]{$C_D$}{throttle discharge coefficient}
\nomenclature[A,1]{$S$}{matter conversion $( mol.m^{-3}.s^{-1} )$ }
\nomenclature[A,1]{$k$}{permeability coefficient in the membrane}
\nomenclature[A,1]{$k_{i,j}$}{nozzle orifice coefficient for i $\in$ $\{sm,em\}$ and j $\in$ $\{in,out\}$ $(kg.Pa^{-1}.s^{-1})$} 
\nomenclature[A,1]{$L_{gc}$}{cumulated length of the gas channel $(m)$}
\nomenclature[A,1]{$U$}{voltage $(V)$}
\nomenclature[A,1]{$W_{gc}$}{width of the gas channel $(m)$}
\nomenclature[A,1]{$W$}{mass flow rate $(kg.s^{-1} / mol.s^{-1} )$}
\nomenclature[A,1]{$S_{h}$}{Sherwood number}
\nomenclature[A,1]{$h$}{convective-conductive mass transfer coefficient $(m.s^{-1})$}
\nomenclature[A,1]{$S_{vl}$}{phase transfer rate of condensation and evaporation $(mol.m^{-3}.s^{-1})$}
\nomenclature[A,1]{$E^{0}$}{standard-state reversible voltage $(V)$}
\nomenclature[A,1]{$x_{v}$}{mole fraction of vapor}
\nomenclature[A,1]{$f_{v}$}{water volume fraction of the membrane}
\nomenclature[A,1]{$f_{drop}$}{liquid water induced voltage drop function}
\nomenclature[A,1]{$V$}{molar volume $( m^{3}.mol^{-1} )$}
\nomenclature[A,1]{$V_{sm}/V_{em}$}{manifold volume $( m^{3} )$}
\nomenclature[A,1]{$K$}{permeability $(m^{2})$}
\nomenclature[A,1]{$K_p/K_d$}{proportionality/derivative constant of the back pressure valve controller $(m^{2}.s^{-1}.Pa^{-1} / m^{2}.Pa^{-1})$}
\nomenclature[A,1]{$D_{i/j}$}{binary diffusivity of two species i and j in open space $(m^{2}.s^{-1})$}
\nomenclature[A,1]{$A_{act}$}{active area $(m^{2})$}
\nomenclature[A,1]{$A_{T}$}{exhaust manifold throttle area $(m^{2})$}
\nomenclature[A,1]{$y_{O_{2}}$}{molar fraction of $O_{2}$ in dry air}
\nomenclature[A,1]{$S_{a}/S_{c}$}{stoichiometric ratio at the anode/cathode}
\nomenclature[A,1]{$R_{e}/R_{p}$}{electron/proton conduction resistance $(\Omega.m^{2})$}
\nomenclature[A,1]{$R$}{universal gas constant $(J.mol^{-1}.K^{-1})$}
\nomenclature[A,1]{$E_{act}$}{activation energy $(J.mol^{-1})$}
\nomenclature[A,2]{\(\kappa_c\)}{overpotential correction exponent}
\nomenclature[A,2]{\(\kappa_{co}\)}{crossover correction coefficient}
\nomenclature[A,2]{\(\alpha_{c}\)}{charge-transfer coefficient of the cathode}
\nomenclature[A,2]{\(\eta\)}{overpotential $(V)$}
\nomenclature[A,2]{\(\Phi\)}{relative humidity}
\nomenclature[A,2]{\(\gamma_{cond}/\gamma_{evap}\)}{overall condensation/evaporation rate constant for water $(s^{-1} / Pa^{-1}.s^{-1})$}
\nomenclature[A,2]{\(\tau\)}{pore structure coefficient}
\nomenclature[A,2]{\(\tau_{cp}/\tau_{hum}\)}{air compressor/humidifier time constant $(s)$}
\nomenclature[A,2]{$\texttt{e}$}{capillary exponent}
\nomenclature[A,2]{\(\sigma\)}{surface tension of liquid water $(N.m^{-1})$}
\nomenclature[A,2]{\(\nu_{l}\)}{liquid water kinematic viscosity $(m^{2}.s^{-1})$}
\nomenclature[A,2]{\(\theta_{c}\)}{contact angle of GDL for liquid water (°)}
\nomenclature[A,2]{\(\gamma_{sorp}\)}{sorption rate $( s^{-1} )$ }
\nomenclature[A,2]{\(\gamma_{H_2}/\gamma_a\)}{heat capacity ratio of $H_2$ and dry air}
\nomenclature[A,2]{\(\lambda\)}{water content}
\nomenclature[A,2]{\(\varepsilon\)}{porosity}
\nomenclature[A,2]{\(\varepsilon_{mc}\)}{volume fraction of ionomer in the CLs}
\nomenclature[A,2]{\(\varepsilon_{p}\)}{percolation threshold porosity}
\nomenclature[A,2]{\(\varepsilon_{c}\)}{compression ratio}
\nomenclature[A,2]{$\texttt{s}$}{liquid water saturation}
\nomenclature[A,2]{$\texttt{s}_{lim}$}{limit liquid water saturation coefficient}
\nomenclature[A,2]{\(\rho\)}{density $( kg.m^{-3} )$}

\nomenclature[B,1]{$\bm{\imath}$}{unit vector along the x-axis}
\nomenclature[B,1]{$K_{shape}$}{shape mathematical factor}
\nomenclature[B,2]{\(a_{s_{lim}}, b_{s_{lim}}, a_{switch}, \texttt{s}_{switch}\)}{fitted values for $f_{drop}$}
\nomenclature[B,2]{\(\alpha, \beta_{1}, \beta_{2}\)}{fitted values for $K_{0}$}
\nomenclature[B,3]{\(\bm{\nabla}\)}{gradient notation}

\nomenclature[C]{\(H_{2}\)}{dihydrogen}
\nomenclature[C]{\(O_{2}\)}{dioxygen}
\nomenclature[C]{\(N_{2}\)}{dinitrogen}
\nomenclature[C]{\(v\)}{vapor}
\nomenclature[C]{\(vl\)}{vapor to liquid}
\nomenclature[C]{\(l\)}{liquid}
\nomenclature[C]{\(w\)}{water}
\nomenclature[C]{\(mem\)}{membrane}
\nomenclature[C]{\(sorp\)}{sorption}
\nomenclature[C]{\(eq\)}{equilibrium}
\nomenclature[C]{\(p\)}{production}
\nomenclature[C]{\(fc\)}{fuel cell}
\nomenclature[C]{\(sat\)}{saturated}
\nomenclature[C]{\(eff\)}{effective}
\nomenclature[C]{\(ref\)}{referenced}
\nomenclature[C]{\(a\)}{anode}
\nomenclature[C]{\(c\)}{cathode}
\nomenclature[C]{\(co\)}{crossover}
\nomenclature[C]{\(cp\)}{compressor}
\nomenclature[C]{\(in\)}{inlet}
\nomenclature[C]{\(out\)}{outlet}
\nomenclature[C]{\(asm\)}{anode supply manifold}
\nomenclature[C]{\(aem\)}{anode exhaust manifold}
\nomenclature[C]{\(csm\)}{cathode supply manifold}
\nomenclature[C]{\(cem\)}{cathode exhaust manifold}

\nomenclature[D]{\(cl/CL\)}{catalyst layer}
\nomenclature[D]{\(acl/ACL\)}{anode catalyst layer}
\nomenclature[D]{\(ccl/CCL\)}{cathode catalyst layer}
\nomenclature[D]{\(gc/GC\)}{gas channel}
\nomenclature[D]{\(agc/AGC\)}{anode gas channel}
\nomenclature[D]{\(cgc/CGC\)}{cathode gas channel}
\nomenclature[D]{\(gdl/GDL\)}{gas diffusion layer}
\nomenclature[D]{\(agdl/AGDL\)}{anode gas diffusion layer}
\nomenclature[D]{\(cgdl/CGDL\)}{cathode gas diffusion layer}
\nomenclature[D]{\(EOD\)}{electro-osmotic drag}
\nomenclature[D]{\(PEMFC\)}{proton exchange membrane fuel cell}

\printnomenclature

\bibliographystyle{elsarticle-num}
\bibliography{PEMFC_nodal_model}
\biboptions{sort&compress}

\end{document}